\documentclass[a4paper,fleqn,usenatbib]{mnras}

\usepackage{newtxtext,newtxmath}

\usepackage[T1]{fontenc}
\usepackage{ae,aecompl}

\usepackage{tabularx}
\usepackage[normalem]{ulem}
\usepackage{multirow}
\usepackage{graphicx}	
\usepackage{amsmath}	
\usepackage{amssymb}	
\usepackage[multiple]{footmisc} 
\usepackage{epstopdf}

\newcommand{\RN}[1]{%
	\textup{\uppercase\expandafter{\romannumeral#1}}%
}

\title[Eclipses of PSR J1810+1744]{The Low-Frequency Radio Eclipses of the Black Widow Pulsar J1810+1744}

\author[E. J. Polzin et al.]{
E.J. Polzin,$^{1}$\thanks{E-mail: elliott.polzin@manchester.ac.uk (EJP)}
R.P. Breton,$^{1}$
A.O. Clarke,$^{1}$
V.I. Kondratiev,$^{2,3}$
B.W. Stappers,$^{1}$
\newauthor
J.W.T. Hessels,$^{2,4}$
C.G. Bassa,$^{2}$
J.W. Broderick,$^{2}$
J.-M. Grie{\ss}meier,$^{5,6}$
C. Sobey,$^{7,8}$
\newauthor
S. ter Veen,$^{2}$
J. van Leeuwen,$^{2,4}$
P. Weltevrede$^{1}$
\\
$^{1}$Jodrell Bank Centre for Astrophysics, School of Physics and Astronomy, The University of Manchester, Manchester M13 9PL, UK\\
$^{2}$ASTRON, the Netherlands Institute for Radio Astronomy, Postbus 2, 7990 AA, Dwingeloo, The Netherlands\\
$^{3}$Astro Space Centre, Lebedev Physical Institute, Russian Academy of Sciences, Profsoyuznaya Str. 84/32, Moscow 117997, Russia\\
$^{4}$Anton Pannekoek Institute for Astronomy, University of Amsterdam, Science Park 904, 1098 XH Amsterdam, The Netherlands\\
$^{5}$LPC2E - Universit{\'e} d'Orl{\'e}ans / CNRS, 45071 Orl{\'e}ans cedex 2, France\\
$^{6}$Station de Radioastronomie de Nan\c{c}ay, Observatoire de Paris, PSL Research University, CNRS, Univ. Orl\'{e}ans, 18330 Nan\c{c}ay, France\\
$^{7}$International Centre for Radio Astronomy Research - Curtin University, GPO Box U1987, Perth, WA 6845, Australia\\
$^{8}$CSIRO Astronomy and Space Science, 26 Dick Perry Avenue, Kensington, WA 6151, Australia\\
}

\date{Accepted XXX. Received YYY; in original form ZZZ}

\pubyear{2017}

\begin{document}
\label{firstpage}
\pagerange{\pageref{firstpage}--\pageref{lastpage}}
\maketitle

\begin{abstract}
We have observed and analysed the eclipses of the black widow pulsar J1810+1744 at low radio frequencies. Using LOw-Frequency ARray (LOFAR) and Westerbork Synthesis Radio Telescope observations between 2011--2015 we have measured variations in flux density, dispersion measure and scattering around eclipses. High-time-resolution, simultaneous beamformed and interferometric imaging LOFAR observations show concurrent disappearance of pulsations and total flux from the source during the eclipses, with a $3\sigma$ upper limit of 36~mJy ($<10\%$ of the pulsar's averaged out-of-eclipse flux density). The dispersion measure variations are highly asymmetric, suggesting a tail of material swept back due to orbital motion. The egress deviations are variable on timescales shorter than the 3.6~hr orbital period and are indicative of a clumpy medium. Additional pulse broadening detected during egress is typically $<20\%$ of the pulsar's spin period, showing no evidence of scattering the pulses beyond detectability in the beamformed data. The eclipses, lasting $\sim13\%$ of the orbit at 149~MHz, are shown to be frequency-dependent with total duration scaling as $\propto\nu^{-0.41\pm0.03}$. The results are discussed in the context of the physical parameters of the system, and an examination of eclipse mechanisms reveals cyclotron-synchrotron absorption as the most likely primary cause, although non-linear scattering mechanisms cannot be quantitatively ruled out. The inferred mass loss rate is a similar order-of-magnitude to the mean rate required to fully evaporate the companion in a Hubble time.
\end{abstract}

\begin{keywords}
pulsars: individual: PSR J1810+1744 -- binaries: eclipsing -- stars: mass-loss -- scattering -- plasmas
\end{keywords}



\section{Introduction}
Black widow pulsars are those which reside in a short orbital period ($\lesssim1$~day) binary system with a low-mass companion star. Similarly to redback pulsars, they differ from typical pulsar -- white dwarf binaries due to the irradiation of the companion star by the pulsar's high-energy wind. Black widows and redbacks represent two classes of these irradiating binaries that are separated by the companion star's mass, with black widow companions typically falling in the range $\sim0.01$--$0.05 M_\odot$, while larger redback companions typically have masses in the range $\sim0.1$--$0.5 M_\odot$ \citep{r13}.\\
The recent surge in the discovery of black widow pulsars from targeted searches of $\gamma$-ray sources found with the \textit{Fermi} Gamma-ray Space Telescope \citep[e.g.][]{rap+12,ckr+15,cck+16} offers an invaluable opportunity to characterise the general population of these puzzling systems. The first such system to be discovered, PSR B1957+20 \citep{fst88}, was found to exhibit eclipses of the radio pulses for $\sim10\%$ of the orbit, centred near inferior conjunction of the companion star. The duration of the eclipses shows that the eclipsing medium must reside outside of the companion's Roche lobe, and optical observations imply a heavily irradiated degenerate companion star \citep{f+88}, suggesting that ablation of the companion star from the pulsar wind constantly replenishes a plasma surrounding the star. Potentially magnetised, it is this plasma that is assumed to cause eclipsing of the pulsar through as-yet undetermined mechanisms \citep[see][for a review]{t+94}. This ablation of the companion is theorised to eventually fully evaporate the star, and as such we are observing the formation route towards isolated Galactic millisecond pulsars \citep{rst89rud}. However, modelling of the winds driven from the companion star in PSR B1957+20 casted doubts on the mass loss being sufficient to achieve complete evaporation within Hubble time \citep{el88}.\\
Until recently, only one other eclipsing black widow pulsar was known in the Galactic field, PSR J2051$-$0827 \citep{s+96}. Observed eclipsing phenomena similar to those seen in PSR B1957+20 revealed the necessity of studies at radio frequencies to act as a unique probe into the pulsar wind, eclipse mechanisms and evolution of black widow systems. However, this limited sample has thus far allowed in-depth radio frequency studies to be carried out for only three Galactic field black widows: PSRs B1957+20 \citep{fbb+90,rt91}, J2051$-$0827 \citep{sbl+01} and J1544+4937 \citep{brr+13}. Thus, many details of these systems remain unclear, and attempts to constrain their global properties require a much larger sample to succeed \citep[e.g.][]{whv+17}.\\
Many features of black widow systems lend themselves to low-frequency observing. Many of the known millisecond pulsars are steep spectrum, i.e. significantly brighter toward lower radio frequency \citep{k+99,kl01,kvh+16}. Increased dispersion and scattering of pulsations typically observed near eclipses in black widows \citep[e.g.][]{sbl+01} is much more prominent at low observing frequencies. In addition, with telescopes such as the LOw-Frequency ARray \citep[LOFAR;][]{v+13,vs10} offering unprecedented sensitivity at frequencies below 200~MHz -- a relatively untouched area for eclipse observations -- valuable opportunities to study these systems are available. A detailed review of observing pulsars with LOFAR is provided in \citet{s+11}.\\
Part of the fresh influx of black widow systems \citep{r13}, PSR J1810+1744 was discovered in a 350~MHz survey of unidentified \textit{Fermi} sources with the Green Bank Telescope \citep{hrm+11}. The 1.66~ms pulsar hosts a companion in a tight, 3.56~hr, orbit and at low radio frequencies it is one of the brightest known millisecond pulsars \citep{kvh+16,kvl+15}. Evidence for irradiation of the companion is found in optical observations showing the signature of a heated inner-face of the tidally locked star \citep{bvr+13,sh14}. Both of these papers cite difficulty in fitting realistic parameters to the observed optical light-curves, and suggest a minimum companion mass, $M_{\text{C}} \gtrsim 0.045 M_\odot$; large for typical black widow companions. \citet{gmr+13} observed the system in X-rays and found no clear evidence for orbital modulation of the X-ray brightness, which is normally ascribed to an intrabinary shock.\\
Here we present multiple low frequency observations of the eclipses of PSR J1810+1744, utilising both beamformed and interferometric methods, in order to probe the mass loss and eclipse mechanisms. Details of the observations are provided in Section~\ref{obs}, with analysis techniques following in Section~\ref{analysis}. The observed eclipses are characterised in Section~\ref{eclipse}, and the measured parameters are discussed in the context of eclipse mechanisms in Section~\ref{mechanism}.

\section{Observations}\label{obs}
Aside from a single observation made with the Westerbork Synthesis Radio Telescope \citep[WSRT;][]{bh74}, all of the observations presented in this paper were carried out using LOFAR over a range of different projects between 2012 December and 2015 February. As listed in Table~\ref{Table: observations}, in total we present $\sim14$~hours of LOFAR data covering two full eclipses, three eclipse ingresses and five eclipse egresses, including one full eclipse and one egress covered in an interferometric observing mode. Both interferometric and beamforming modes were used in order to provide image plane and high time resolution observations, respectively.\\ 
\begin{table*}
	\centering
	\caption{List of observations. $^a$L260707, L260713, L260719, L260725, L260731, L260737, L260743 and L260749. $^b$Eight consecutive observations of 30 minutes, separated by 7 minute intervals. $^c$LOFAR tied-array beams. $^d$Centre frequency of observation.}
	\label{Table: observations}
	\begin{tabular}{lcccccr}
		\hline	
		Date & Telescope & Project ID & ObsID & Duration & Orbital phase & Comments \\
		\hline
		$2015 / 02 / 19$ & LOFAR & LC2\_039 & $^a$ & $8\times30$m & 0.20--1.55$^b$ & Interferometric\\
		& & & & & & \& beamformed\\
		$2014 / 09 / 19$ & LOFAR & LC2\_026 & L243355 & 1h & 0.24--0.52 & 2 TABs$^c$\\
		$2014 / 09 / 18$ & LOFAR & LC2\_026 & L243365 & 1h & 0.21--0.49 & 2 TABs$^c$\\
		$2014 / 06 / 17$ & LOFAR & LC2\_026 & L231759 & 1h & 0.25--0.53 & 2 TABs$^c$\\
		$2013 / 10 / 02$ & LOFAR & LC0\_011 & L181912 & 5m & 0.18--0.20 & \\
		$2013 / 07 / 26$ & LOFAR & DDT\_005 & L166106 & 2h 6m & 0.70--1.29 &\\
		$2013 / 07 / 25$ & LOFAR & DDT\_005 & L166110 & 2h 42m & 0.81--1.56 &\\
		$2013 / 07 / 24$ & LOFAR & DDT\_005 & L165450 & 2h 12m & 0.20--0.83 &\\
		$2012 / 12 / 20$ & LOFAR & LC0\_011 & L81280 & 20m & 0.12--0.22 &\\
		$2011 / 06 / 06$ & WSRT & S11A008 & 11102762 & 5h & 0.60--2.00 & 345~MHz$^d$\\
		\hline
	\end{tabular}
\end{table*}
\subsection{LOFAR beamformed observations}
\begin{figure}
	\includegraphics[width=\columnwidth]{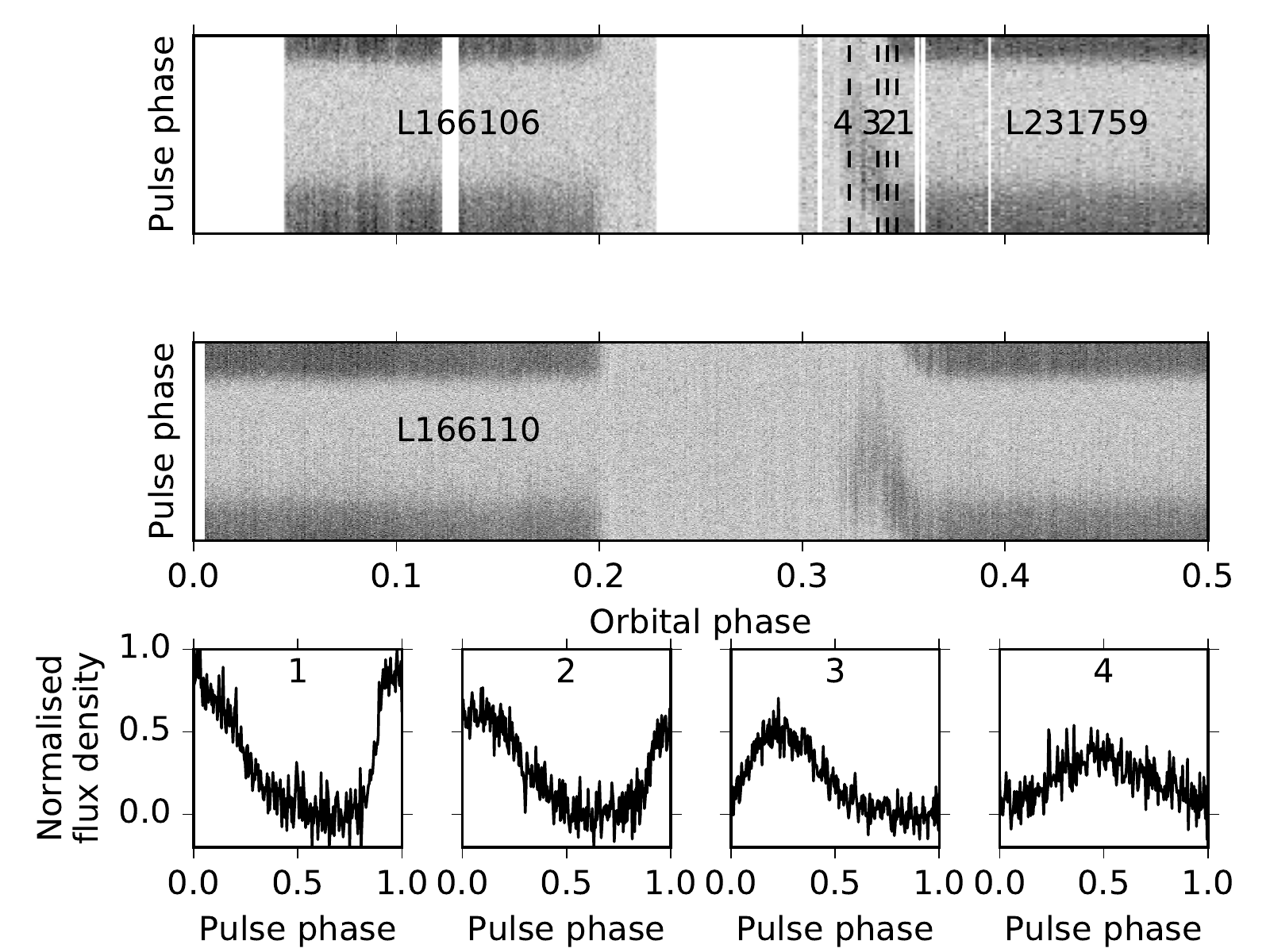}
	\caption{Calibrated, folded beamformed data from observations of PSR J1810+1744. Top / Middle: Pulse phase against orbital phase for three observations near eclipse. Observation IDs are indicated in the plots. An orbital phase of 0.25 corresponds to inferior conjunction of the companion. Delays in the pulse arrival times are apparent in the two eclipse egresses. Bottom: Pulse profiles, each integrated over 1~min of data from the observation L231759. The orbital phases of the centre of each 1~min integration are shown as dashed vertical lines in the upper-right egress plot, with numbers corresponding to the labels in the profile plots. Each profile has been normalised to the maximum flux density of profile no.1.}
	\label{fig:BF_data}
\end{figure}

These observations utilised the wide fractional bandwidth of the LOFAR High Band Antennas (HBA), spanning the frequency range 110--188~MHz at a centre frequency of 149~MHz. Dual orthogonal linear polarisation signals from 18--23 LOFAR Core stations were directed into the Central Processing Facility (CEP) correlator where the station data streams were coherently summed.\\
The majority of the data were collected using the LOFAR coherent Stokes mode to form 1--2 Tied-Array Beams (TABs) from the Core stations. Observations with two TABs had one beam centred on the coordinates of the pulsar and a second beam displaced by $\sim 6$~arcmin in declination. The second, off-source beam provided a means of discriminating between loss of pulsations and loss of total flux from the pulsar, and is described further in Section~\ref{eclipse}. Pre-processing of these data was performed using the LOFAR Pulsar Pipeline (PulP). PulP \citet{ahm+10,s+11} is an offline pipeline that utilises the software packages \texttt{PRESTO}\footnote{https://github.com/scottransom/presto} \citep{r01}, \texttt{DSPSR}\footnote{http://dspsr.sourceforge.net/} \citep{vb11} and \texttt{PSRCHIVE}\footnote{https://psrchive.sourceforge.net/} \citep{hvm04} to form dedispersed, folded data products for analysis. Specifically here, PulP was used for coherent dedispersion and folding (\texttt{DSPSR}) of the data into archive files with 5~second sub-integrations, 256 pulse phase bins and 400 frequency channels, each of 195.3~kHz width, and RFI excision with \texttt{PSRCHIVE}'s \texttt{paz}.\\
Further processing of the data consisted of both flux and polarisation calibration. Polarisation calibration followed the method of \citet{nsk+15} to apply the inverse of the instrumental response using the Jones matrices generated from the Hamaker-Carozzi beam model \citep[][with \texttt{mscorpol}\footnote{https://github.com/2baOrNot2ba/mscorpol}]{h06} for each beam pointing and frequency channel. Application of the Jones matrices to perform the calibration used \texttt{PSRCHIVE}'s \texttt{pac} command. \texttt{PSRCHIVE}'s \texttt{pam} was then used to convert the calibrated data into the four Stokes parameters, $I$, $Q$, $U$, $V$.\\
Flux calibration was performed using \texttt{lofar\_fluxcal.py} from the \texttt{LOFAR-BF-pulsar-scripts} package\footnote{https://github.com/vkond/LOFAR-BF-pulsar-scripts}, detailed in \citet{kvh+16}, to correct for instrumental gain variations as a function of time and frequency. In brief, the script scales each sub-integration, frequency channel and polarisation based on a theoretical flux density estimation using the signal-to-noise of the data and assumed values for instrumental effects. Specifically, this once again makes use of the Hamaker-Carozzi beam model, system temperature estimations using the \citet{hss+82} sky model scaled to LOFAR frequencies as $\nu^{-2.55}$ \citep{lmo+87} and model antenna temperatures from \citet{wv11} as a function of frequency. Also accounted for are the fraction of operational HBA tiles, fraction of channels and sub-integrations that were excised due to RFI during PulP, and a power law scaling of coherent station summation based on the number of HBA stations used. The noise level in the data is estimated using the mean and standard deviation of a pre-defined off-pulse region of the profile in each sub-integration, channel and polarisation. In order to avoid biassing this estimation, it was necessary to dedisperse each sub-integration individually (at the eclipse edges in particular) so that the pulsar flux did not smear into the pre-defined off-pulse region. The importance of flux calibration was clearest in the long-duration observations, where variations in beam shape and shadowing of antenna tiles became more pronounced over the range of beam elevations. The calibrated data were re-folded into sub-integrations of duration 5--60~seconds depending on the signal-to-noise in each observation. Examples of the calibrated beamformed data are shown in Fig.~\ref{fig:BF_data}, where the radio pulses can be seen to be delayed and reduced in flux density near the eclipses.
\subsection{WSRT beamformed observation}
WSRT data from 2011 June were used to provide a comparatively higher-radio-frequency probe into the eclipse. These data were in the form of a single 5~hour observation using the PuMa-\RN{2} pulsar backend \citep{ksv08} with a bandwidth between 310--380~MHz, centred on 345~MHz. Using \texttt{DSPSR}, the data were coherently dedispersed and folded into 1~min sub-integrations, 128 pulse phase bins and 512 frequency channels, each of 156~kHz width. Here, a simpler flux calibration method was used which, although not providing an absolute flux density scale, normalised the profile in each sub-integration and frequency channel based on the off-pulse noise level in order to remove telescope gain variations to first-order. Once again the data were dedispersed separately for each sub-integration around eclipse orbital phases to avoid leakage of pulsar flux into the baseline noise estimation.\\
\subsection{Timing ephemerides}
Initial visual inspection of the data showed the pulses to drift significantly in phase within each observation, highlighting the lack of a satisfactory long-term timing solution for this pulsar. As often observed in black widows \citep[e.g][]{aft94,svf+16}, due to as-yet undetermined mechanisms, the orbital period can vary on relatively short timescales, leading to difficulties in defining an ephemeris that can be used to correctly fold data over a range of observations. To account for this, \texttt{TEMPO2}\footnote{https://sourceforge.net/projects/tempo2/} \citep{hem06} was used to find a satisfactory ephemeris for each individual observation by fitting the out-of-eclipse pulse times of arrival (TOAs) through adjustment of the spin frequency and its derivative (F0, F1), binary period (PB) and time of ascending node (T0) timing parameters. See Appendix~\ref{Sec: orb_ephem} for orbital ephemerides used.\\
\subsection{LOFAR interferometric observations}\label{obs_interf}
\begin{figure*}
	\includegraphics[width=\textwidth]{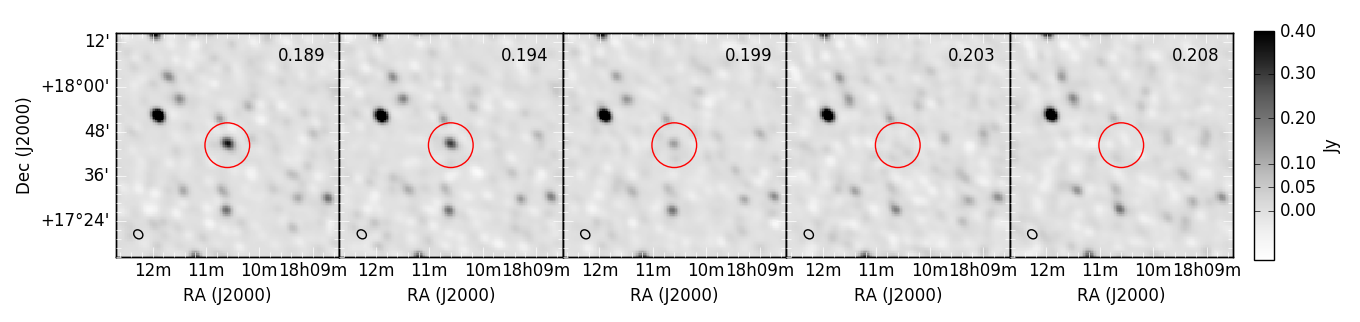}
	\caption{A sequence of LOFAR images showing eclipse ingress. The position of PSR J1810+1744 is at the centre of the red circle in each image. The images were formed with an integration time of 1~min, and are cropped to a $1^{\circ}\times1^{\circ}$ square centred on the pulsar coordinates. The orbital phase at the centre point of each integration time is shown in the top right hand corner of each image, with a phase of 0.25 corresponding to inferior conjunction of the companion. The synthesised beam shape is shown in the lower left hand corner of each image.}
	\label{fig:images}
\end{figure*}

Simultaneous interferometric and beamformed LOFAR observations were undertaken in 2015 February. Using the $\sim 4^{\circ}$ full-width at half-maximum of the LOFAR Core HBA beam (at $\sim150$~MHz) the interferometric data were collected with spectral and temporal resolution of $\sim 3$~kHz and 1~second, respectively. The maximum baseline of the LOFAR Core stations is $\sim2$~km, corresponding to a resolution of $\sim165$~arcseconds at 150~MHz. These raw visibilities were reduced in an averaging pipeline and the resulting output stored on the LOFAR Long Term Archive (LTA) in measurement sets with time intervals of 10~seconds duration and frequency channels of $\sim 49$~kHz width. Further details of the LOFAR pipeline are available in \citet{hmp+10}. The 5~hour observing time was split into alternate beam pointings of 7~min centred on flux calibrator 3C295, followed by 30~min centred on the pulsar, leading to coverage of slightly over one orbit.\\
The measurement sets were processed using standard LOFAR methods, with \texttt{AOFLAGGER} \citep{odb+10,ovr12,odz12} used for RFI flagging and Black Board Self-calibration \citep[\texttt{BBS};][]{pvd+09} used for both flux and phase calibration. Flux calibration of 3C295 was carried out using a pre-defined LOFAR sky model for the source\footnote{https://github.com/lofar-astron/prefactor/tree/master/skymodels}\footnote{\citet{sh12} quote $\sim3\%$ uncertainty on flux density at 149~MHz.} \citep{sh12} and the resulting gain solutions were applied to the pulsar measurement sets. We performed self-calibration using a sky model of the source region from the TGSS LOFAR Sky Model Creator\footnote{http://tgssadr.strw.leidenuniv.nl/doku.php}. Imaging of the calibrated visibilities was carried out using \texttt{CASA}\footnote{https://casa.nrao.edu/}, thus no primary beam correction was applied. This lack of correction is justified due to the location of the pulsar at the centre of the image, thus the effect of a changing beam over time is negligible. Due to the brightness of PSR J1810+1744, we were able to make images of the field using 1~min time intervals of data, over the full $\sim 80$~MHz bandwidth, with the pulsar detectable at $\sim40\times$ the typical RMS noise of $\sim12$~mJy in the 1~min images. A sequence of the 1~min images, zoomed into a $1^{\circ}\times1^{\circ}$ area centred on the pulsar, is shown in Fig.~\ref{fig:images}, as the pulsar enters into eclipse.\\
\section{Analysis}\label{analysis}
\subsection{Analysis of beamformed data}\label{analysis_bf}
\begin{figure*}
	\includegraphics[width=0.98\textwidth]{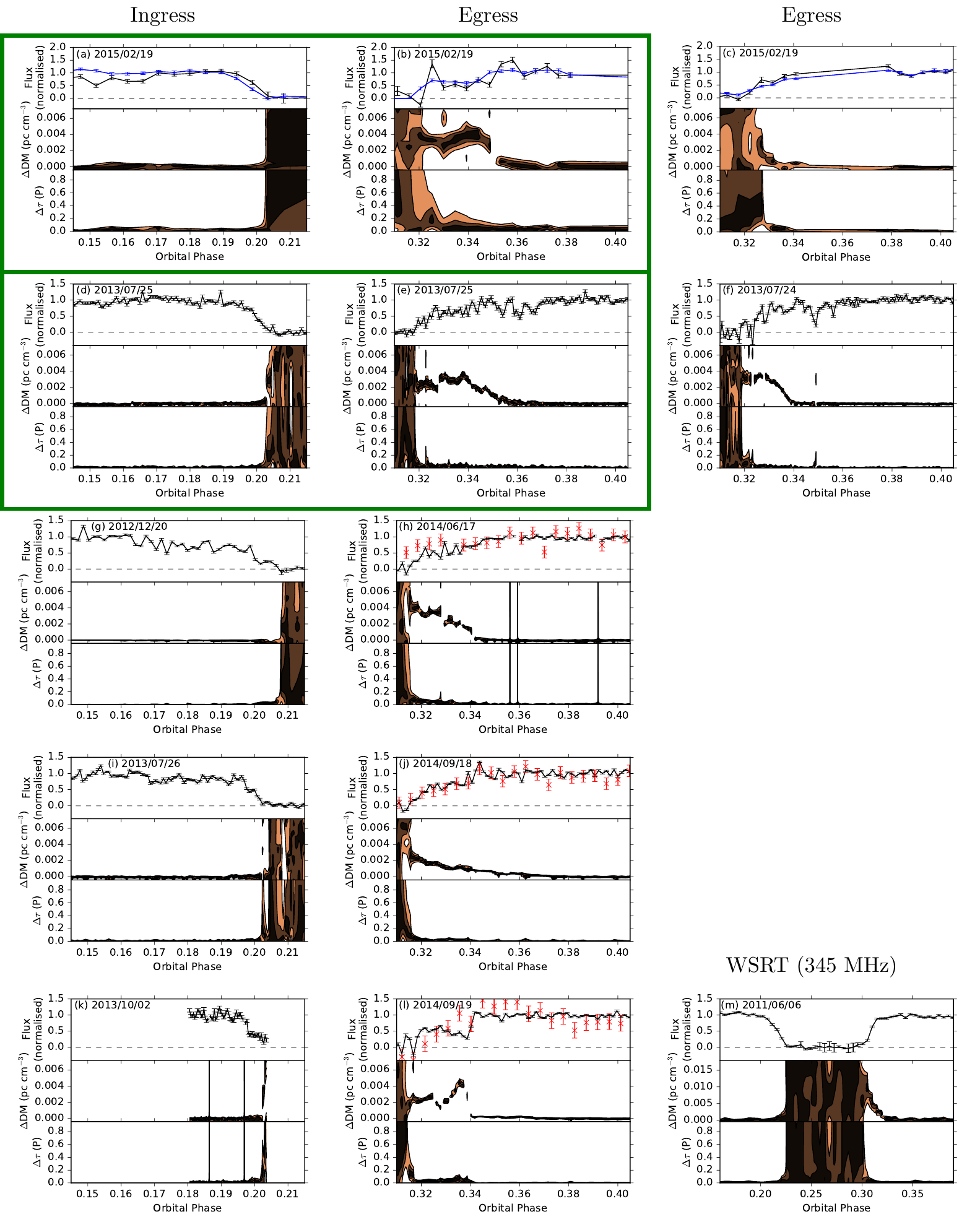}
	\caption{Detected flux density (top panel of each sub-plot (a)--(m)), deviation of dispersion measure, $\Delta$DM (middle panels) and deviation of scattering timescale, $\Delta\tau$ (bottom panels) relative to the out-of-eclipse values. The out-of-eclipse flux density is normalised to unity for each observation independently, with black curves showing beamformed flux density, blue curves showing interferometric flux density ((a)--(c)), and red crosses showing beamformed flux density normalised by an off-source beam ((h), (j), (l)). For $\Delta$DM and $\Delta\tau$ panels the black, dark and light brown contours represent 1, 2 and 3$\sigma$ uncertainties, respectively. An orbital phase of 0.25 corresponds to the companion's inferior conjunction. Sub-plot (m) corresponds to the 345~MHz WSRT observation and is plotted on a different phase and $\Delta$DM scale for display purposes, while the rest show 149~MHz LOFAR data. Green boxes represent ingress and egress from a single eclipse.}
	\label{fig:params_vs_time}
\end{figure*}

Key to our understanding of eclipsing pulsars is detecting variations in pulsar flux density, dispersion measure (DM) and scattering throughout the orbit in order to constrain the physical parameters of the material in the eclipsing medium. Here we adopted a template fitting method, written in \texttt{Python}\footnote{https://www.python.org/}, to measure these parameters in our data as a function of orbital phase.\\
Initially, two pulsar templates were generated: one for LOFAR, and one for WSRT. The templates were made by spin-phase-aligning and summing together in time all of the out-of-eclipse observations of the pulsar to ensure maximum signal-to-noise while avoiding any pulse smearing near eclipse edges. To aid in breaking degeneracies between small changes in DM and scattering, the templates were two-dimensional, with frequency and pulse phase binning equal to that of the data. A third-order Savitzky-Golay filter \citep{sg64} was applied along both axes to smooth out the small scale noise variations that remained. An array of further templates was made by dispersing and scattering the stock, out-of-eclipse templates with given DM deviations, $\Delta$DM, and scattering timescale deviations, $\Delta\tau$, over ranges of $\Delta$DM and $\Delta\tau$ pre-determined by inspection of the data. For LOFAR, templates were made for $0 \leq \Delta$DM~$\leq 0.01$ in $5\times10^{-5}$~pc~cm$^{-3}$ steps, and $0 \leq \Delta\tau \leq P$, where $P$ is the pulse period, in steps of $0.02P$. For WSRT, templates were made for $0 \leq \Delta$DM~$\leq 0.03$ in $1\times10^{-4}$~pc~cm$^{-3}$ steps, and similarly $0 \leq \Delta\tau \leq P$, in steps of $0.02P$. We assumed validity of the cold plasma dispersion relation, $\Delta t$\textsubscript{DM}~$\propto \nu^{-2}$, and modelled scattering as a convolution of the pulse with a frequency dependent exponential, $\frac{1}{\tau} \exp^{-t/\tau}$, where $\tau \propto \nu^{-4}$ \citep{l71,lj76}.\\
For each sub-integration of data a least-squares fit of the array of templates was performed, returning a minimised $\chi^2$ value per template. Both the baseline and scale factor of each template were free parameters in the fits, treating the template as a single entity (as opposed to each frequency channel individually) assuming that scintillation will not affect the spectral index over time, as discussed in Section~\ref{scintillation}.  Thus, for each sub-integration of data, a two-dimensional $\chi^2$ map was obtained over the ranges of $\Delta$DM and $\Delta\tau$. Finding the maximum likelihood of $\Delta\tau$ allowed the distribution of $\chi^2$ values for $\Delta$DM to be plotted as a function of sub-integration of data, and vice-versa for $\Delta\tau$. The resulting contour maps are shown in Fig.~\ref{fig:params_vs_time} with contours plotted for $1\sigma$, $2\sigma$ and $3\sigma$ deviations from the minimum $\chi^2$, taking into account any degeneracy between $\Delta$DM and $\Delta\tau$. Fig.~\ref{fig:params_vs_time} also shows the scale factor, and corresponding uncertainty, of the best fitting template for each sub-integration, directly relating to the detected \textit{pulsed} flux density from the pulsar throughout the orbit. For display, these light-curves are normalised so that the out-of-eclipse flux densities are equal to unity. The average out-of-eclipse flux density that we measure for the calibrated beamformed data is $297\pm33$~mJy, however the nominal uncertainty here is assumed to be an underestimate of the true value, and instead we propose a $50\%$ uncertainty of $\sim150$~mJy based on previous studies of LOFAR flux density measurements \citep{bkk+16,kvh+16}.\\
\subsubsection{Modification of the model}
To investigate the validity of the scattering model used, we re-created sets of templates for different frequency power-law exponents, $\tau \propto \nu^{-\alpha}$, with $\alpha = 2.0$, 2.5, 3.0, and 3.5, based on recent work by \citet{gkk+17} whereby forward fitting of simulated scattering profiles to LOFAR observations yielded values of $1.5 \leq \alpha \leq 4.0$. In addition, templates were made assuming the scattering model of \citet{w72} for a thick screen positioned close to the pulsar, with scattering function,
\begin{equation}
\left(\frac{\pi\tau}{4t^3}\right)^{1/2} \exp^{-\pi^2 \tau/16t}.
\end{equation}
However, due to both the low-level scattering variations present and the low signal-to-noise of the pulse profile on timescales short enough to track variations, no statistically significant deviations from the presented results were found.\\
Alternatively, to further investigate the dispersion we referred to the general form of the plasma dispersion law \citep{pw92} where, as discussed in \citet{hsh+12}, the quartic term,
\begin{equation}
\Delta t_{\text{EM}} = \frac{\text{EM}}{4\nu^4},
\end{equation}
may become detectable for low observing frequencies traversing a clumpy medium. Here, the Emission Measure, EM~$=\int_{0}^{D}n_{\text{e}}^2\text{d}l$, where $n_{\text{e}}$ (cm$^{-3}$) represents number density of electrons and $D$ (pc) is the distance to the source. Incorporating this additional term into our fits yielded some qualitatively interesting initial results, with occasional `spikes' during egress reaching $\text{EM}\sim10^5$~pc~cm$^{-6}$. In contrast, compact H\RN{2} regions typically have $\text{EM}\gtrsim10^7$~pc~cm$^{-6}$ \citep[Table 1 of][]{k02}. However, statistical comparison between this, and the original simpler DM model, using both the Akaike and Bayesian Information Criterion \citep[][respectively]{a74,s78} suggested that the likelihood of the more complex EM model is slightly lower than that of the more favourable simple DM model. This likelihood takes into account the fact that the increased number of degrees of freedom in the EM model is expected to lower the minimum $\chi^2$ of the fit, irrespective of the model being a better or worse representation of the "true" mechanism. In light of these inconclusive results, we plan to further investigate the general plasma dispersion law for low-frequency observations of black widow pulsars.
\subsection{Analysis of interferometric data}
Interferometric data are sensitive to both pulsed flux and any continuum, un-pulsed flux from the black widow system. Each of the 1~min images, described in Section~\ref{obs_interf}, was analysed with \texttt{PyBDSF}\footnote{http://www.astron.nl/citt/pybdsm/} to measure the average flux density from the pulsar within that interval. \texttt{PyBDSF} automatically identifies "islands" of emission in the image, and for each island it fits two-dimensional Gaussians to individual sources. The flux density of each source is calculated directly from the fitted Gaussians, thus with knowledge of the expected coordinates of the pulsar, the flux density and associated uncertainty were retrieved. For the case of eclipsing pulsars this process can falter as the signal-to-noise decreases, or fully disappears, near eclipse. To counter this we used a high signal-to-noise "detection image", made by combining several out-of-eclipse observations, to identify islands of emission and consequently passed these onto the 1~min images for flux density extraction. The resulting light-curve is shown plotted with the simultaneously observed beamformed flux density as blue points in Fig.~\ref{fig:params_vs_time}(a), (b) and (c).\\
Out-of-eclipse, we detected PSR J1810+1744 at an average flux density of $483\pm5$~mJy in the 78~MHz wide LOFAR band centred on 149~MHz. This is significantly higher than the value quoted in Section~\ref{analysis_bf}, even with the estimated $50\%$ uncertainty on that measurement. Although this discrepancy can suggest systematic errors in the flux density calibration methods, we note that the interferometric flux density quoted here is measured from observations covering just one orbit of the system, whereas the beamformed flux density from Section~\ref{analysis_bf} is averaged over observations spanning $>2$~years and thus is much more susceptible to the effects of refractive scintillation (see Section~\ref{scintillation}).\\
Direct comparison with the 150~MHz flux density found in the GMRT TGSS survey \citep{ijm+17} required us to re-image the field with the same $\sim17$~MHz bandwidth used in that survey, resulting in a measured flux density for PSR J1810+1744 of $380\pm10$~mJy from our observations. Investigation of the MJDs of the observations used to calculate the flux density quoted from TGSS led to the finding that the pulsar was in typically eclipsed orbital phases for $\sim30\%$ of the scans, thus scaling to account for this leads to a TGSS flux density from PSR J1810+1744 of $390\pm40$~mJy; consistent with the value found here. Finally, a similar analysis was carried out by re-imaging the field with an 8~MHz bandwidth centred on 150~MHz in order to allow direct comparison with the equivalent flux density found in the GLEAM survey \citep{hch+17}. This led us to find a flux density of $335\pm15$~mJy, consistent with the flux density of $321\pm88$~mJy reported from GLEAM.
\subsection{Scintillation}\label{scintillation}
Studying flux density variations in eclipsing pulsars can be complicated by interstellar scintillation effects that can mask, or mimic, genuine flux density variations in the system depending on the bandwidth and timescale of the scintles \citep{fbb+90,sbl+01}. At LOFAR frequencies however, the decorrelation bandwidth for diffractive scintillation for PSR J1810+1744 is $\sim 2$~kHz, thus any effects are averaged out over the $80$~MHz band \citep{akh+14}. Scaling this to 345~MHz using the measured dependence of decorrelation bandwidth on observing frequency, $\Delta\nu_{\text{d}} \propto \nu^{4.5}$, for this pulsar \citep{akh+14} gives a decorrelation bandwidth for diffractive scintillation of $\sim 90$~kHz for the WSRT observation, thus also too small to have significant effects over the full band. Conversely, refractive scintillation can affect the observed flux densities, however with a decorrelation timescale on the order of weeks it has little effect on flux density variations within orbital timescales. As such, the \textit{short timescale} flux density variations presented here are assumed to be largely independent of interstellar scintillation. On the other hand, we caution that the measured out-of-eclipse, absolute flux densities in this paper are expected to be influenced by the long-term refractive variability. Measurements of the flux density of the source in each of the beamformed LOFAR observations used here appear to show a smooth trend over the $\sim 2$~year range, with a factor of $\sim 2$ separating the lowest flux density from the highest.
\section{Eclipse characteristics}\label{eclipse}
\subsection{Individual eclipses}
Fig.~\ref{fig:params_vs_time} shows the measured deviations of flux density, DM and scattering time for all detected eclipses. Here we highlight the variability between separate eclipses and the timescales over which these occur.
\subsubsection{Scattering}
Leading into eclipse, and during eclipse recovery, where the flux density $\gtrsim20\%$ of the out-of-eclipse level, little to no additional scattering is seen. However, very close to eclipse edges the results from the fits are unreliable. Here, inspection of the observed data showed no evidence of increased scattering tails in the pulsations and any apparent sharp increases (e.g. Fig.~\ref{fig:params_vs_time}(k)) are assumed to be artefacts of the fit to low signal-to-noise data. The eclipse egresses in general show $\Delta\tau <$~20\% of the pulse period, with occasional short timescale rises up to $\sim40\%$. This low level scattering is far from that required to reduce the pulsed flux density beyond detection, unless a sharp rise were to occur at eclipse boundaries. Such a sharp rise would have to occur on a timescale shorter than 10~seconds, equivalent to the shortest integration times presented here, to avoid detection. In addition, in the higher-frequency WSRT observation in Fig.~\ref{fig:params_vs_time}(m) the pulsar flux remains detectable for orbital phases closer to inferior conjunction of the companion, thus probing further into the eclipsing medium, and shows no evidence of steep scattering variations.
\subsubsection{Dispersion measure}
Similarly to scattering, the DM (or electron column density) shows no clear evidence of increasing prior to loss of pulsations in eclipse ingress. Significant deviations from the out-of-eclipse electron column density can regularly be seen in eclipse egresses. The duration of these deviations varies from $\sim 5$--$10$~min (2--5~$\%$ of the orbit) between observations. Features in the electron column density profiles show clear distinctions in each eclipse, with notable extremes in Fig.~\ref{fig:params_vs_time}(l) -- where multiple rises and falls are present, including a sharp boundary to the out-of-eclipse level, and Fig.~\ref{fig:params_vs_time}(j) -- where the electron column density decreases slowly and smoothly away from eclipse. Of particular interest for the timescale of variations is the fact that these two eclipses were observed on consecutive days. Placing even tighter constraints on the time variability of the DM features during egress are the consecutive eclipses in Fig.~\ref{fig:params_vs_time}(b) and (c). The prominent DM feature in Fig.~\ref{fig:params_vs_time}(b), extending out to orbital phase 0.36, shows no sign of detection in the egress observed one orbit earlier in Fig.~\ref{fig:params_vs_time}(c), demonstrating that the DM features are variable within the orbital timescale of 3.6~hours.\\
Assuming these electron column density deviations are a result of material within, or in close proximity to, the binary system we are likely to be sampling the outer edges of the eclipsing medium with such low observing frequencies being sensitive to small variations in DM. The extended egress variations are indicative of a `tail' of material, swept-back due to orbital motion \citep{fbb+90,sbl+01,whv+17}. In this case, our results suggest that the material in the tail is dynamic within orbital timescales, and sharp electron column density variations imply the material to be clumpy in nature. There is some evidence of a commonly appearing peak, or flattening, of the electron column density around orbital phases 0.33--0.34, which could be suggestive of a quasi-static dense region, or represent the geometry of the tail.\\
At 149~MHz the maximum detected $\Delta\text{DM}_{149}<0.006$~pc~cm$^{-3}$, while at 345~MHz, $\Delta\text{DM}_{345}<0.015$~pc~cm$^{-3}$. For dispersion smearing across the width of a channel to reach the pulse period, and thus remove the pulses, would require $\Delta\text{DM}_{149}>1.3$~pc~cm$^{-3}$ and $\Delta\text{DM}_{345}>38.3$~pc~cm$^{-3}$; many orders of magnitude above those detected. However, in the case of short timescale DM variations, an additional case of DM smearing can be encountered. Should the DM change significantly over the duration of a single sub-integration, the resulting pulse, averaged over the integration time, would be smeared by an amount dependent on the observed frequency. Due to the measurement of DM utilising templates with the full frequency resolution of the data, this smearing would be problematic if the change in DM within a sub-integration, $\Delta\left(\Delta\text{DM}\right)$, became large enough as to cause the integrated pulse within a channel to smear over the full pulse phase. This would have the most significant effect on the channels with the lowest frequencies, and would occur at the bottom of the respective frequency bands for $\Delta\left(\Delta\text{DM}_{149}\right)>0.005$~pc~cm$^{-3}$ and $\Delta\left(\Delta\text{DM}_{345}\right)>0.04$~pc~cm$^{-3}$ for 149~MHz and 345~MHz, respectively. Although not an issue at 345~MHz, at 149~MHz the maximum gradient detected across a single sub-integration is $\sim0.002$~pc~cm$^{-3}$, which would smear the pulse by $0.7$~ms, or $\sim40\%$ of the pulse period at the lowest side of the frequency band, causing an apparent decrease in detected flux density within the sub-integration. In general however, this gradient is much lower than that required to smear the pulses beyond detection.
\subsubsection{Flux density}
At 149~MHz, the pulsed flux density typically decreases much faster at eclipse ingress in comparison to the egress, in agreement with that expected for a swept-back tail of material. However, in the single 345~MHz eclipse this relationship is much more symmetric, although with a hint of the relationship being reversed where pulsed flux density recovers marginally quicker post-eclipse, being largely independent of the material in the extended tail.\\
In all observed eclipse ingresses, and most notably that in Fig.~\ref{fig:params_vs_time}(g), the pulsed flux density begins to attenuate without any detected increase in the electron column density. Should the attenuation be a result of additional material along the line-of-sight, then a $\Delta\text{DM}_{149}<0.0003$~pc~cm$^{-3}$ would be required to avoid detection amongst the typical out-of-eclipse $3\sigma$ uncertainties. This would correspond to an electron column density of eclipsing material, $N_{\text{e}} \lesssim 10^{15}$~cm$^{-2}$.\\
Similarly to the DM, at 149~MHz the pulsed flux density takes $\sim5$--$10$~min to recover in eclipse egress. The egress in Fig.~\ref{fig:params_vs_time}(l) shows a clear anti-correlation between the flux density recovery and DM, with the pulsed flux density rising and falling with corresponding troughs and peaks in DM, respectively. The flux density variation associated with the sharp DM boundary to the out-of-eclipse level should be interpreted with care however, as a bias could be introduced by DM smearing with this steep gradient. In contrast to this, the pulsed flux density recovery in other eclipse egresses show no obvious correlation with the electron column density, as multiple peaks and troughs in pulsed flux density occur with no detected change in DM. In addition, the electron column density often shows a plateau, or even peak, closely after eclipse, throughout which the pulsed flux density continuously recovers, apparently independently of the electron column density. Although the pulsed flux density varies significantly in the egress region, there are no detectable mini-eclipses \citep[such as those seen in, e.g., PSR B1744$-$24A and PSR J1023+0038;][respectively]{ljm+90,asr+09} in any of the observations outside of the orbital phases 0.15--0.40 shown here. Also note that any apparent re-appearances of flux density during an eclipse (e.g. Fig.~\ref{fig:params_vs_time}(d)) show no sign of being realistic upon inspection of the data, and thus are likely to be artefacts of the fitting method.\\
For the three eclipse egress observations that utilised a second, simultaneous off-source beam we were able to investigate the variation in the un-pulsed continuum flux density from the pulsar, independent of smearing or scattering of the pulse. By assuming both beams are affected equally by the telescope, sky and RFI, the off-source beam was used to normalise, and essentially flat-field, the pulsar centred beam. Thus, by averaging the normalised data over all pulse phases and subtracting unity, the remaining flux density in each sub-integration is proportional to that of the total flux density from the pulsar, independent of pulsations. The resulting light-curves, plotted in red on Fig.~\ref{fig:params_vs_time}(h), (j) and (l), appear to show the total flux density disappearing in eclipse.\\
Stronger evidence of this is provided by the simultaneous beamformed and interferometric observations shown in Fig.~\ref{fig:params_vs_time}(a), (b) and (c). The interferometric flux density, sensitive to the total, pulse phase averaged flux density from the pulsar, is shown to track the same disappearance and re-appearance as the pulsed flux, giving clear evidence that flux is removed from the line-of-sight, rather than smeared or scattered.
\subsection{Global properties}
\begin{figure}
	\includegraphics[width=\columnwidth]{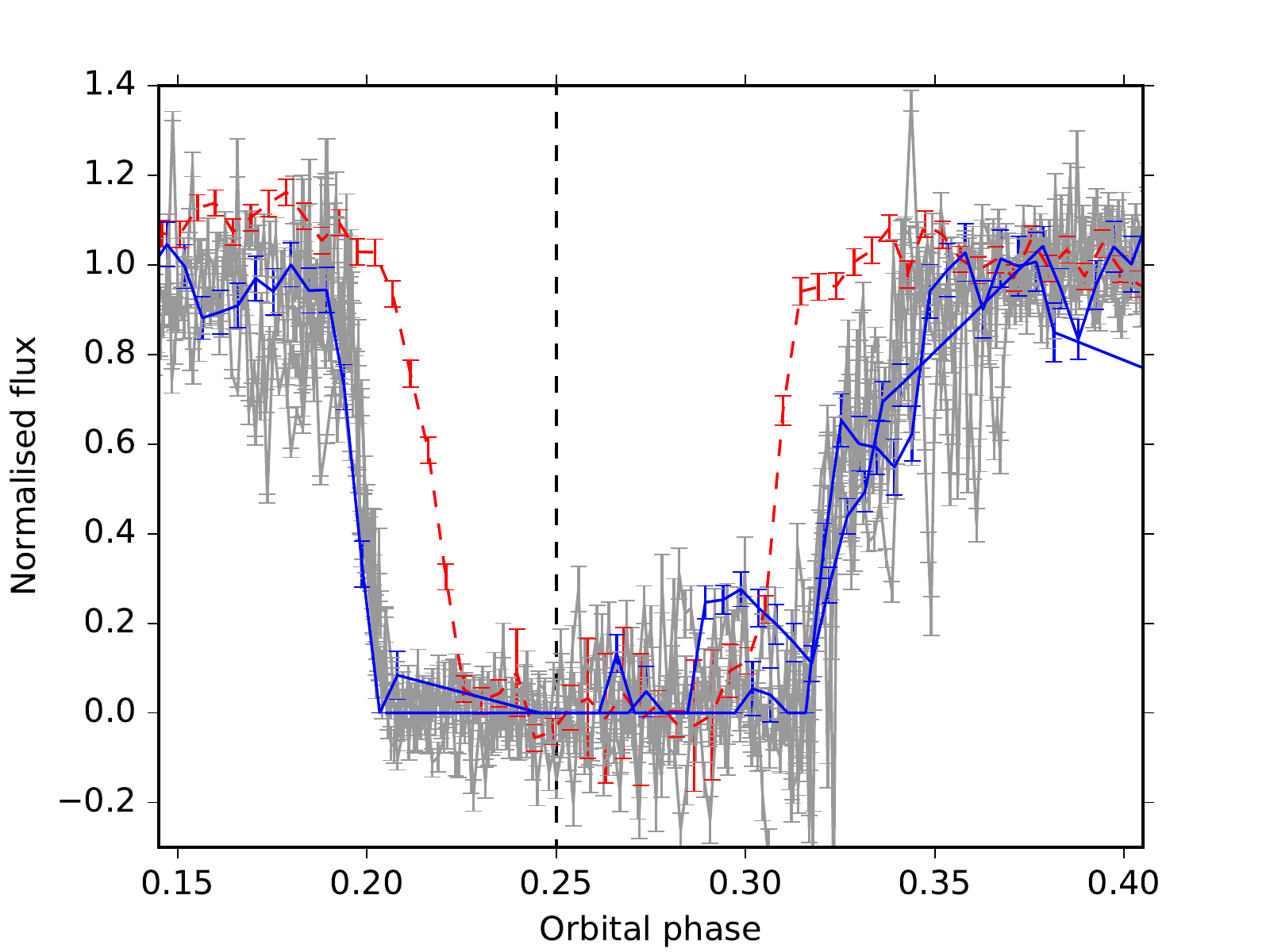}
	\caption{Measured flux density as a function of orbital phase for all observations. The out-of-eclipse flux density is normalised to unity. The pulsed flux density at 345 and 149~MHz is shown in red and grey, respectively. The total continuum flux density at 149~MHz is shown in blue. The dashed line at phase 0.25 corresponds to inferior conjunction of the companion star.}
	\label{fig:flux_all}
\end{figure}

\begin{figure}
	\includegraphics[width=\columnwidth]{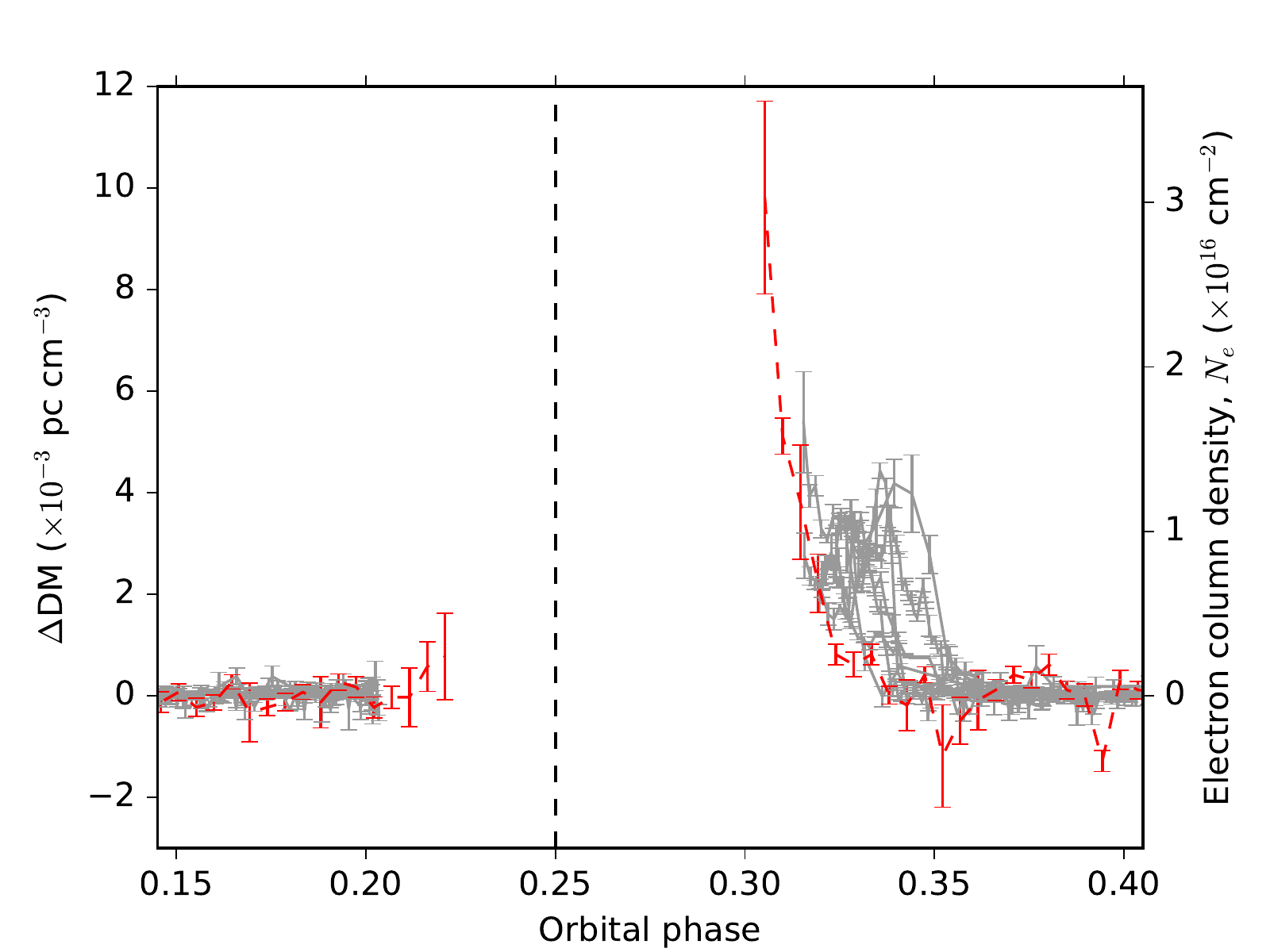}
	\caption{Deviation of dispersion measure relative to mean out-of-eclipse value for all beamformed observations. The observations at 345 and 149~MHz are shown in red and grey, respectively. The dashed line at phase 0.25 corresponds to inferior conjunction of the companion star.}
	\label{fig:dm_all}
\end{figure}

\begin{figure*}
	\includegraphics[width=\textwidth]{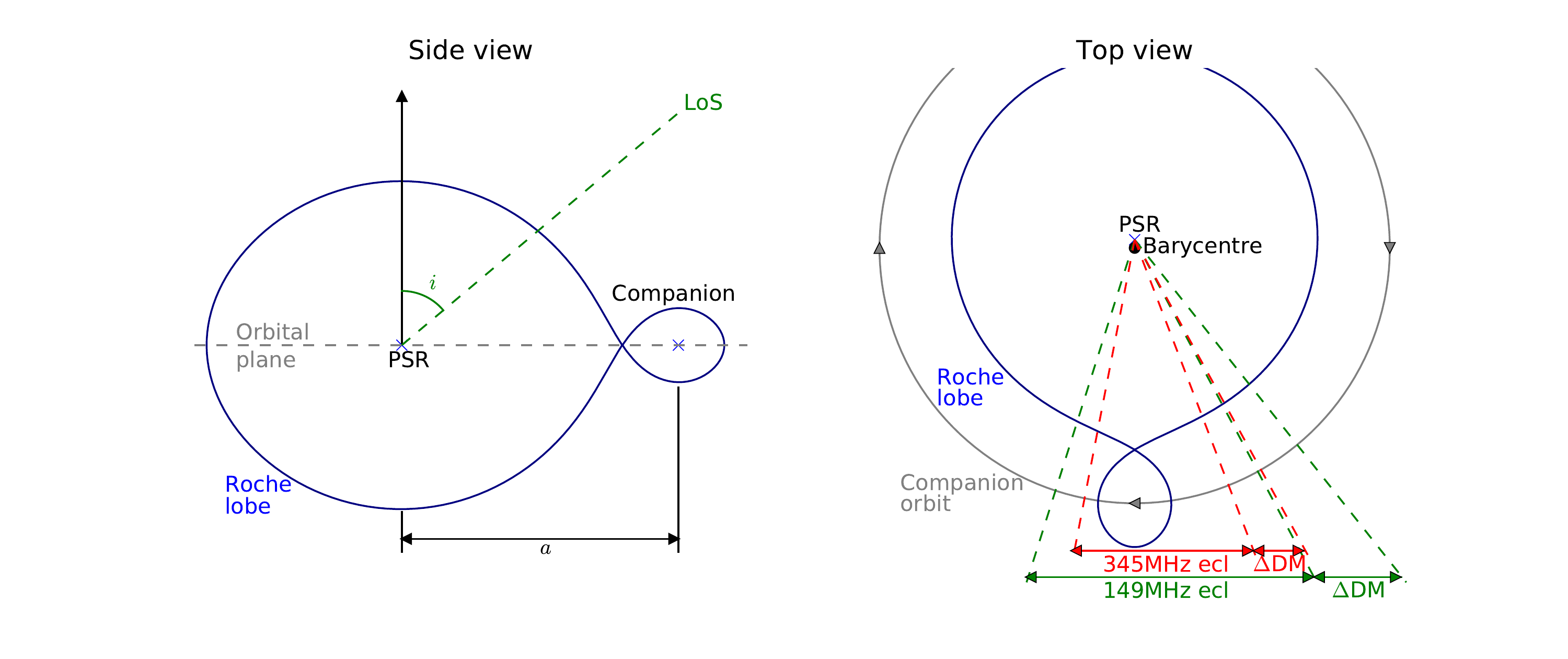}
	\caption{Expected geometry of the PSR J1810+1744 system, schematic is drawn to-scale. The Roche lobe and companion orbit are calculated assuming a pulsar mass of 1.4$M_\odot$, companion mass of 0.045$M_\odot$ and an orbital separation, $a=1.33R_\odot$ \citep{bvr+13}. Note that no attempt is made to display the physical size of the companion star. Left: a `side-on' view from the orbital plane, the line-of-sight (LoS) is drawn assuming an orbital inclination, $i=50^{\circ}$ \citep{bvr+13,sh14}. Right: a view from perpendicular to the orbital plane, the line-of-sights corresponding to eclipse edges (half-maximum flux density) and post-eclipse egress DM variations are shown by red (345~MHz) and green (149~MHz) dashed lines. The grey arrows indicate the companion's direction of motion.}
	\label{fig:geometry}
\end{figure*}

Figs.~\ref{fig:flux_all} and \ref{fig:dm_all} show the flux density and DM variations for all combined observations, respectively. The mean duration, radius, centre points and asymmetry of the eclipses are shown in Table~\ref{tab:eclipse}. The duration is taken to be the full-width at half-maximum of the flux density, and was calculated by fitting the normalised ingress and egress flux densities with Fermi-Dirac functions, $f=1/\left(e^{\frac{\phi+p_1}{p_2}}+1\right)$, where $f$ is the normalised flux density, $\phi$ is orbital phase and $p_1$ and $p_2$ are the fitted free parameters. Calculation of the eclipse radius, $R_{\text{E}}$, assumes the eclipsing medium resides at the orbit of the companion, with orbital separation $a = 1.33 R_\odot$ and mass ratio $q=0.045/1.4$ \citep{bvr+13}. The asymmetry of the eclipses is characterised by the ratio of egress to ingress phases about $\phi = 0.25$, i.e. $(\phi_{\text{eg}}-0.25)/(0.25-\phi_{\text{in}})$, where $\phi_{\text{in}}$ and $\phi_{\text{eg}}$ correspond to the phase at half-maximum flux density found from the Fermi-Dirac fits for ingress and egress, respectively.\\
Similarly to black widow PSRs B1957+20 \citep{fbb+90}, J2051$-$0827 \citep{s+96} and J1544+4937 \citep{brr+13}, the low frequency eclipse lasts for $\sim 10\%$ of the orbit, irrespective of the distinctly different system parameters. The centre points of the eclipses, $\phi_{\text{c}}$, occur at a similar phase after inferior conjunction of the companion at both 149~MHz and 345~MHz, in contrast to that seen for PSR B1957+20 \citep{fbb+90} where the eclipse is centred near orbital phase 0.25, and PSR J1544+4937 \citep{brr+13} where the eclipse is centred slightly prior to 0.25. Due to the shorter duration of the 345~MHz eclipse the asymmetry, as defined by the ratio of egress to ingress durations, is unusually larger than at 149~MHz. Although the electron column density can be highly variable from one eclipse to the next, the disappearance and re-appearance of flux can be seen to occur regularly at the same orbital phases, with very little deviation.\\
Note that these calculations assume the orbit to be edge-on. Fits to optical light-curves for this system in both \citet{bvr+13} and \citet{sh14} find best fitting orbital inclinations of $\sim50^{\circ}$, suggesting that we are sampling the outer edges of the medium where the density is likely to be lower and eclipse width shorter than that in the orbital plane. The expected geometry of this system is shown in Fig.~\ref{fig:geometry}. Should the eclipsing material be spherically symmetric and centred in the orbital plane at the distance of the companion, then an edge-on view of this system would result in eclipse radii of 1.2$R_\odot$ and 1.1$R_\odot$ for 149~MHz and 345~MHz, respectively; similar in magnitude to the orbital separation.\\
Using the polarisation calibrated LOFAR data, we measure the pulsar to have an average rotation measure, RM~$= 90.48\pm0.02$~rad~m$^{-2}$, prior to ionospheric correction (see Sobey et al., in prep. for further information, including the ionosphere--corrected value). Using the measured RMs to correct for Faraday rotation, we find average linear and circular polarisation fractions of $\sim0.01$ and $\sim0.05$, respectively. The small polarisation fractions did not allow us to place any constraints on the magnetic fields in the eclipse medium.\\
\begin{table*}
	\centering
	\caption{Duration, $\Delta\phi_{\text{eclipse}}$, radius, $R_{\text{E}}$, centre point, $\phi_{\text{c}}$, and asymmetry, $\Delta\phi_{\text{eg}}/\Delta\phi_{\text{in}}$, for the 149~MHz and 345~MHz eclipses. $\Delta\phi_{\text{eclipse}}$ and $\phi_{\text{c}}$ are in units of orbital phase. The top row shows $\Delta\phi_{\text{eclipse}}$, $\phi_{\text{c}}$ and $\Delta\phi_{\text{eg}}/\Delta\phi_{\text{in}}$ measured using the half-maximum flux density as eclipse boundaries, with $R_{\text{E}}$ calculated assuming an orbital inclination, $i=90^{\circ}$. The bottom row shows the extrapolated duration and radius of the eclipse \textit{within the orbital plane} assuming that the observed eclipse is for an orbital inclination, $i=50^{\circ}$, the eclipse medium is spherically symmetric and centred in the orbital plane at the distance of the companion. The uncertainties on the extrapolated values are formally calculated through error propagation, however assume zero uncertainty on the orbital inclination.}
	\label{tab:eclipse}
	\begin{tabular}{lccccr}
		\hline
		& $\nu$ (MHz) & $\Delta\phi_{\text{eclipse}}$ & $R_{\text{E}}$ & $\phi_{\text{c}}$ & $\Delta\phi_{\text{eg}}/\Delta\phi_{\text{in}}$\\
		\hline
		\multirow{2}{*}{$i=90^{\circ}$}& 149 & $0.130\pm0.002$ & $(0.51\pm0.01)R_\odot$ & $0.265\pm0.002$ & $1.48\pm0.02$\\
		& 345 & $0.090\pm0.002$ & $(0.37\pm0.01)R_\odot$ & $0.264\pm0.002$ & $1.88\pm0.02$\\
		\hline
		\multirow{2}{*}{$i=50^{\circ}$}& 149 & $0.23\pm0.001$ & $(1.2\pm0.01)R_\odot$ & & \\
		& 345 & $0.21\pm0.001$ & $(1.1\pm0.01)R_\odot$ & & \\
		\hline
	\end{tabular}
\end{table*}
\subsubsection{Material in the system}
The combined $\Delta\text{DM}$ measurements in Fig.~\ref{fig:dm_all} demonstrate the strong asymmetry of material either side of the low frequency eclipse, much the same as that observed for black widow PSR B1957+20 \citep{rt91}. The Roche lobe of the companion, assuming $R_{\text{L}}=0.19R_\odot$ and orbital inclination of $50^{\circ}$ \citep{bvr+13}, is not intersected by our line-of-sight and as such the medium causing these eclipses cannot be gravitationally bound to the companion. Fig.~\ref{fig:geometry} highlights the remarkable extent to which the eclipsing medium may have to be distributed in order to intercept the line-of-sight. Should the inclination estimate be doubted, in order for the line-of-sight to touch the Roche lobe an orbital inclination~$\gtrsim80^{\circ}$ would be required, and even for an edge-on view the Roche lobe would span the orbital phases $\sim 0.23$--$0.27$; inside the eclipse at both ingress and egress. Attempts to constrain the density of material in the eclipsing medium are at best an order-of-magnitude estimates. The asymmetry between ingress and egress, with barely any detectable rise in electron column density at ingress provides very little insight to the leading edge of the medium. In addition, being far from the gravitational grasp of the companion, the eclipse tail material could well extend far outside the orbit.\\
With this in mind, to allow comparison between systems we choose to follow previous considerations where it is assumed that all of the material is contained within an approximately spherical region, centred on the companion, of diameter equal to the eclipse width \citep{t+94}. Thus, at 149~MHz with an eclipse width, $2R_{\text{E}}\sim 1.0R_\odot$ and electron column density, $N_{\text{e}}\approx1\times10^{16}$~cm$^{-2}$, the electron density at egress is
\begin{equation}
n_{\text{e},149} \approx 1.4\times10^5 \left(\frac{2R_{\text{E}}}{1.0R_\odot}\right)^{-1}~\text{cm}^{-3}.
\end{equation}
While at 345~MHz, with eclipse width, $2R_{\text{E}}\sim 0.8R_\odot$ and egress column density, $N_{\text{e}}\approx3\times10^{16}$~cm$^{-2}$, we find
\begin{equation}
n_{\text{e},345} \approx 5\times10^5 \left(\frac{2R_{\text{E}}}{0.8R_\odot}\right)^{-1}~\text{cm}^{-3}.
\end{equation}
At 345~MHz this is an order of magnitude larger than the equivalent density found for PSR B1957+20 in \citet{t+94}, but similar to that found for PSR J2051$-$0827 at 436~MHz \citep{s+96}.\\
As stated in Section 6.3 of \citet{t+94}, should the ablated material become entrained in the pulsar wind then the mass loss rate from the companion can be estimated as, $\dot{M}_{\text{C}} \sim \pi R_{\text{E}}^2m_{\text{p}}n_{\text{e}}V_{\text{W}}$, where $R_{\text{E}}$ is taken to be the eclipse half-width and represents the radius of the projected circle through which the mass is assumed to be lost, and $V_{\text{W}}$ is the outflow velocity of ablated material in the pulsar wind. Taking into account the expected geometry of the PSR J1810+1744 system, we take as $R_{\text{E}}$ the inferred radius of eclipse material should it be a sphere centred on the companion, i.e. 1.2$R_\odot$ and 1.1$R_\odot$ for 149~MHz and 345~MHz, respectively. If the momentum flux of the ablated material is taken to be equal to the momentum flux of the pulsar wind at the distance of the companion, then $V_{\text{W}} = (U_{\text{E}}/n_{\text{e}}m_{\text{p}})^{1/2}$. Assuming the pulsar wind to be isotropic then we find the energy density of the wind at the companion distance,
\begin{equation}
U_{\text{E}} = \frac{\dot{E}}{4\pi c a^2} = 12.3\left(\frac{a}{1.33R_\odot}\right)^{-2}~\text{ergs~cm}^{-3},
\end{equation}
where $\dot{E}$ is the spin-down power of the pulsar. Thus, for the 149~MHz observations using $R_{\text{E}} = 1.2R_\odot$ and $n_{\text{e}} = 1.4\times10^5$~cm$^{-3}$, we find an estimated mass loss rate of $\dot{M}_{\text{C}} \sim 6\times10^{-13}~M_\odot$~yr$^{-1}$. Instead using the inferred values from the 345~MHz observation of $R_{\text{E}} = 1.1R_\odot$ and $n_{\text{e}} = 5\times10^5$~cm$^{-3}$, we find $\dot{M}_{\text{C}} \sim 1\times10^{-12}~M_\odot$~yr$^{-1}$. In comparison, to fully evaporate the companion over a timescale of $\sim5$~Gyr would require an average mass loss rate of $\dot{M}_{\text{C}} \sim 9\times10^{-12}~M_\odot$~yr$^{-1}$, only factor of $\sim10$ larger than that estimated at 345~MHz. Considering that these estimates depend rather heavily on the assumptions made about the geometry and mechanics of the mass loss, it may not be unreasonable for the companion to be fully evaporated within Hubble time. However, we note that the long-term orbital dynamics of the system will also likely influence an evolution of the mass loss over time; in the presence of negligible magnetic braking and gravitational radiation, mass loss from the system will act to widen the orbital separation between the two bodies. When combined with spin-down of the pulsar, it would be naively expected that this predicted decrease in irradiation of the companion star over time will reduce the probability of complete evaporation.\\
In this model, with the ablated material entrained in the pulsar wind, we note that the observed low column densities suggest that the material would be emitted approximately radially from the orbit. This assumes that the material initially flows from the companion at the escape velocity, thus the momentum flux of the thin material would be far lower than that of the pulsar wind and it would be carried away at $\sim V_{\text{W}}$. With such a high-velocity radial flow we would not expect to see large asymmetries in the measured column densities about inferior conjunction of the companion, in contrast to that observed. To reconcile this, it can be proposed that the companion hosts a magnetosphere filled with material of much higher density, and that is compact enough so as to not be intersected by our line of sight. This scenario allows for the denser magnetosphere to be swept back due to the orbital motion causing measurable asymmetry of the eclipses. Magnetic reconnection between the pulsar wind magnetic field and the companion's magnetosphere allows for material to leak into the pulsar wind, detectable as additional egress column densities as the companion continues through its orbit (Thompson, priv. comm.).
\subsubsection{Frequency dependence}
\begin{figure}
	\includegraphics[width=\columnwidth]{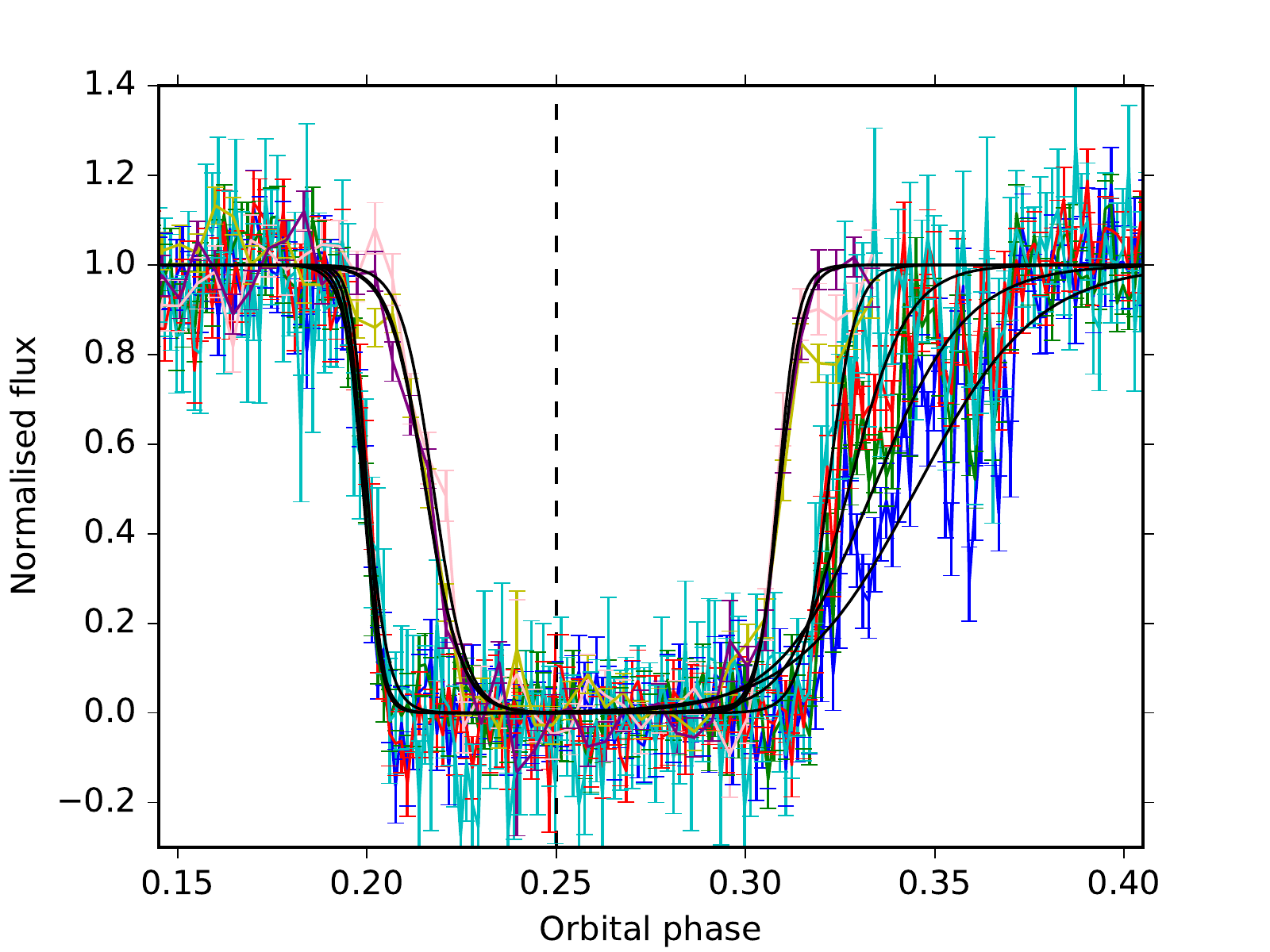}
	\caption{Normalised flux density for the full eclipses observed on 2011 June 6 (WSRT) and 2013 July 25 (LOFAR). The flux densities as measured in each of the frequency sub-bands centred on 118 (blue), 134 (green), 154 (red), 176 (cyan), 321 (pink), 344 (yellow) and 369~MHz (purple) are shown. Wider eclipses are observed for lower frequency sub-bands. The black curves show least squares fits of Fermi-Dirac functions to the ingresses and egresses in each frequency sub-band.}
	\label{fig:ecl_duration}
\end{figure}

\begin{figure}
	\includegraphics[width=\columnwidth]{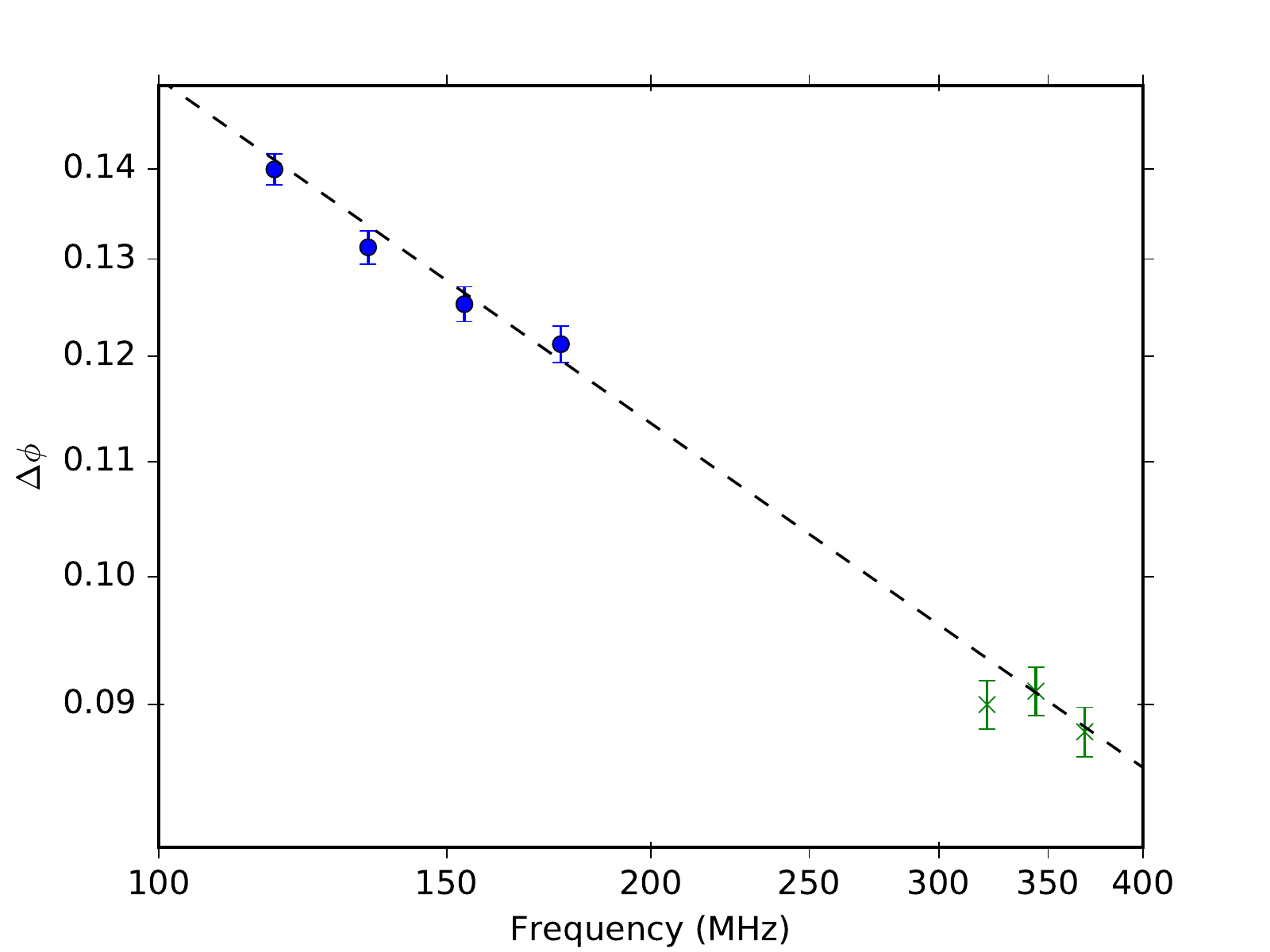}
	\caption{Eclipse duration, $\Delta\phi_{\text{eclipse}}$, taken as full-width at half-maximum of out-of-eclipse flux density, for observed full eclipses. The error bars correspond to $1\sigma$ uncertainties from the Fermi-Dirac fits to the measured flux densities. The dashed line shows a best fit power law with $\alpha = -0.41$. Different colours represent separate eclipses.}
	\label{fig:freq_duration}
\end{figure}

\begin{figure}
	\includegraphics[width=\columnwidth]{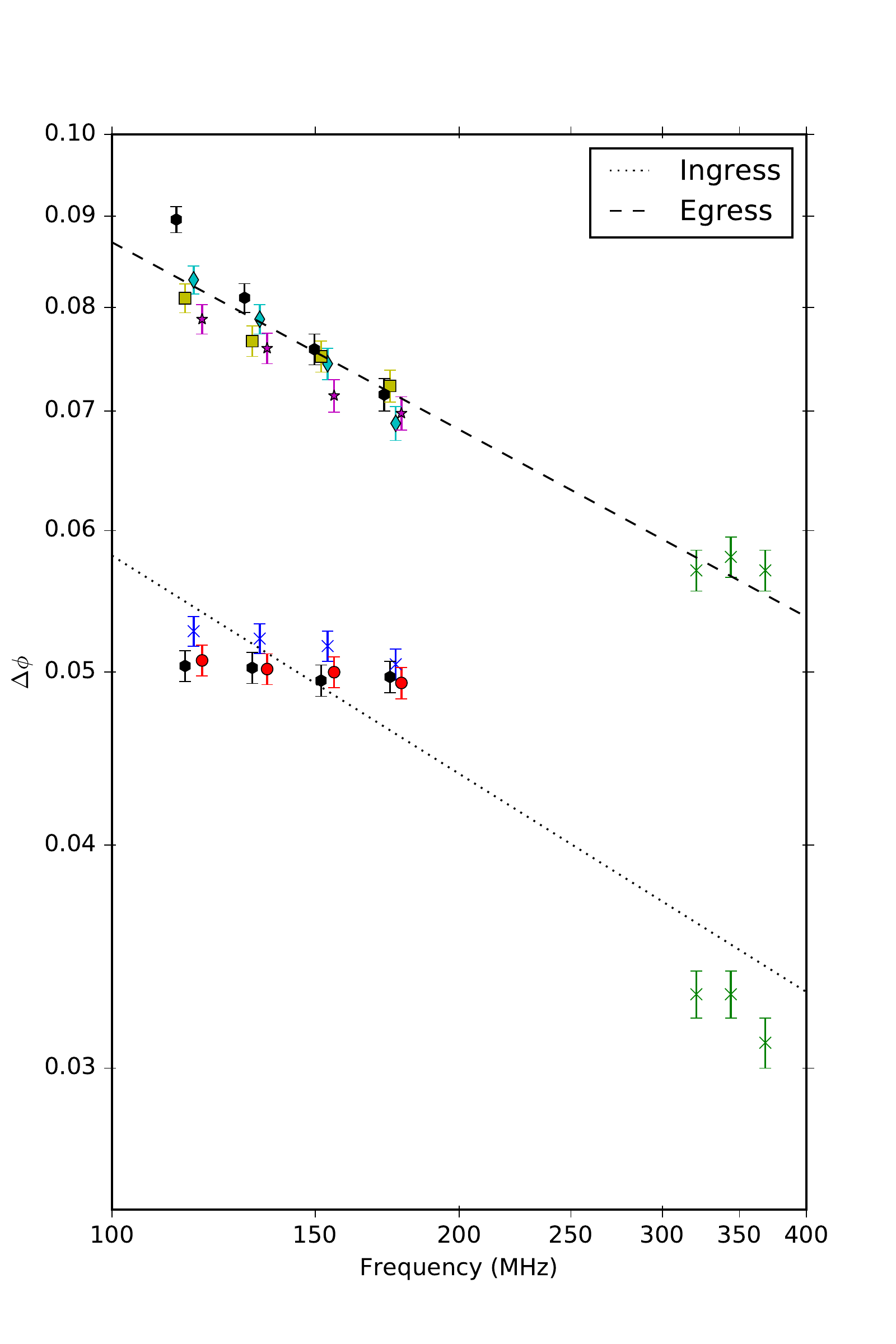}
	\caption{Ingress and egress durations, taken as $\lvert0.25-$~phase of half-maximum flux density$\rvert$. Top: four LOFAR and one WSRT egress observations, the dashed line shows a best fit power law with $\alpha = -0.35$. Bottom: three LOFAR and one WSRT ingress observations, the dotted line shows a best fit power law with $\alpha = -0.41$.  The error bars correspond to $1\sigma$ uncertainties from the Fermi-Dirac fits to the measured flux densities. Different coloured symbols represent separate eclipses. Durations, taken at 118, 134, 154, 176, 321, 344 and 369~MHz, are plotted with small frequency offsets for clarity.}
	\label{fig:freq_egress_ingress}
\end{figure}

The wide fractional bandwidth of these observations allows for a thorough investigation of the frequency dependence of the eclipse across the range of observed frequencies. The LOFAR observations were re-analysed with the bandwidth split into four sub-bands of logarithmically increasing widths, centred on 118, 134, 154 and 176~MHz. Similarly, the WSRT eclipse was re-analysed with three sub-bands centred on 321, 344 and 369~MHz. The eclipse, as observed in each frequency sub-band, is shown in Fig.~\ref{fig:ecl_duration}, along with the corresponding Fermi-Dirac fits plotted as black solid lines. The eclipse durations, $\Delta\phi_{\text{eclipse}}$, were calculated for two of the observed full eclipses, taken as the width of the eclipse at half of the out-of-eclipse flux density, and are shown in Fig.~\ref{fig:freq_duration}. A power law was fit to the data of the form $\Delta\phi_{\text{eclipse}} \propto \nu^\alpha$, where $\nu$ represents observing frequency, finding $\alpha=-0.41\pm0.03$, consistent with $\alpha\sim-0.4$ found for PSR B1957+20 \citep{fbb+90} and redback PSR J2215+5135 \citep{bfb+16}.\\
Due to the observed asymmetry of the eclipses, we carried out similar analysis treating ingress and egress independently. The corresponding durations were taken to be the deviation of half-maximum flux density for each ingress and egress from orbital phase 0.25. Power laws of the same form were fit to both ingress and egress durations, and are shown in Fig.~\ref{fig:freq_egress_ingress}. While the egress durations appear to be well modelled by a best fit power law with $\alpha=-0.35\pm0.05$, the ingress durations do not seem well explained by a single power law. All three of the plotted LOFAR ingresses display little frequency dependence over the LOFAR band, suggesting a much shallower $\alpha$ at these frequencies. This could be expected for a sharp boundary at the leading edge of the eclipse medium, however it would need to be of very specific density in order to allow the continuation of 345~MHz radiation further into the medium. Equally, the 345~MHz DM measurements show no sign of a step change in electron column density at the 149~MHz ingress phase, thus it is not clear what could be causing such a phenomenon.\\
One potential source of this observed effect is the plasma frequency within the medium reaching that of the observing frequency. For the plasma frequency to reach these thresholds would require electron densities of $n_{\text{e}} = 3\times10^8$~cm$^{-3}$ and $n_{\text{e}} = 1.5\times10^9$~cm$^{-3}$ for 149~MHz and 345~MHz, respectively. These are many orders of magnitude higher than those inferred earlier from the observed electron column densities and eclipse widths, and as such could only be a feasible cause of eclipses if the material was contained within a thin region along the line-of-sight. Such a region may occur in an intrabinary shock, where the opposing pulsar and companion winds reach a pressure balance. X-ray observations of this system show possible evidence of count-rate modulation characteristic of such a shock, however due to large measurement uncertainties no significant detection is claimed \citep{gmr+13}.\\
As an added insight into the eclipse mechanism, the line-of-sight optical depth as a function of orbital phase can be calculated for each frequency sub-band. Assuming a power law dependency of optical depth with frequency of the form $\tau \propto \nu^\beta$, and fitting to egress data for four LOFAR observations and the single WSRT observation, we find $\beta$ to increase approximately linearly over orbital phases 0.32--0.34 within the range $-2 \lesssim \beta \lesssim -5$. Similar fits to the eclipse ingresses find $\beta\sim-1$ over the much narrower orbital phase range.\\
\section{Discussion of eclipse mechanisms}\label{mechanism}
Once again, to allow direct comparisons we choose to follow the steps of previous black widow studies to evaluate possible eclipse mechanisms using the analysis in \cite{t+94} as a basis. Immediately here we can rule out pulse smearing and small-angle scattering as eclipse mechanisms due to the disappearance of flux density seen in the interferometric observations. Refraction is commonly disfavoured due to the lack of expected frequency dependence of the resulting eclipses and smaller than required excess delays in pulse arrival times at eclipse boundaries \citep{t+94,sbl+01,brr+13}. Similarly, with corresponding pulse delays $< 1$~ms measured here, to produce a caustic would require such a high density gradient at eclipse edges that the eclipse duration would be almost independent of frequency \citep[Section \RN{3}(b) of][]{el90}, contrary to the clear frequency dependence in duration observed here, making refraction unlikely to be the primary source of the low frequency eclipses in PSR J1810+1744. We note, however, that the apparent clumpy nature of the eclipsing medium, and very shallow frequency dependence for the 100--200~MHz eclipse ingress could allow refraction to have some role in the process. Additionally, refraction would provide an explanation for observed features such as the brief enhancements in flux density seen in the egress in Fig.~\ref{fig:params_vs_time}(j), where the line-of-sight could pass a caustic.\\
For free-free absorption, using Equation (11) of \citet{t+94} with an electron column density from the 345~MHz egress, $N_{\text{e}}\approx3\times10^{16}$~cm$^{-2}$, and absorption length equal to the eclipse width of $5.6\times10^{10}$~cm, we find a required temperature of the eclipse medium, $T\lesssim10^3 f_{\text{cl}}^{2/3}$~K, with clumping factor $f_{\text{cl}}= \left\langle n_{\text{e}}^2 \right\rangle / \left\langle n_{\text{e}} \right\rangle$, for an optical depth $\tau_{\text{ff}}>1$. The same temperature constraint is found for the 149~MHz eclipse parameters. This is equal to the limit calculated for PSR B1957+20 in \citet{t+94} and is understood to be orders of magnitude colder than that expected in such a medium with a realistic clumping factor.\\
Should the eclipse region material increase in density closer to the companion (as expected from the observations), off-axis radio beams could reflect or refract from the higher density regions into the line-of-sight radio beam, causing induced Compton scattering in the line-of-sight radiation. As explained in \citet{t+94} this would alter the observed radiation spectrum, appearing to act as absorption of low-frequency radiation. Using Equation (26) of \citet{t+94} with the measured flux density of PSR J1810+1744, $S_\nu^0 = 480$~mJy, a spectral index $\alpha = -2.3$ \citep{mkb+17}, electron column density $N_{\text{e}}\approx10^{16}$~cm$^{-2}$, a source distance of $d=2.0$~kpc and orbital separation $a = 9.3\times10^{10}$~cm from \citet{bvr+13}, we find an optical depth at 149~MHz of $\tau_{\text{ind}} \lesssim 0.5 M$, where $M\sim(R_{\text{C}}/2r)^2$ is the magnification factor of the reflected radio beam, $R_{\text{C}}$ is the radius of curvature of the reflecting region and $r$ is the distance from the centre of curvature to the scattering region \citep[Section 2.4.1. of][]{t+94}. For PSR J1810+1744 we estimate the radius of curvature of the higher density region to be equal to the Roche lobe radius, $R_{\text{C}} \sim 0.19R_\odot$, and take the scattering region to be at the distance of the 149~MHz ingress from the centre of the companion, assuming the material surrounds the companion and that the system has an orbital inclination of $50^{\circ}$, thus $r\sim0.95R_\odot$. As such, we calculate $M\sim0.01$ and a corresponding optical depth at 149~MHz, $\tau_{\text{ind}} \lesssim 5\times10^{-3}$. This suggests that induced Compton scattering is unlikely to be sufficient to cause the low frequency eclipses.\\
With the relatively high radio flux density of PSR J1810+1744 the non-linear mechanism of a stimulated Raman scattering parametric instability may be significant. Here, the radio flux incident on the eclipsing medium is theorised to generate turbulence in the plasma that can consequently scatter further radio waves out of the line-of-sight. Using distance to the pulsar, $d=2.0$~kpc and $a = 9.3\times10^{10}$~cm, we estimate the radio flux from the pulsar at the companion's orbit to be 0.02~ergs~cm$^{-2}$~s$^{-1}$~Hz$^{-1}$ at 149~MHz. Following the analysis using Equation (32) of \citet{t+94} with an electron density at the 149~MHz eclipse, $n_{\text{e}} \approx 1.4\times10^5$~cm$^{-3}$, we find that the critical flux incident on the eclipsing medium required for strong scattering is exceeded here if the temperature of the medium, $T \gtrsim 10^4$~K, which is expected to be satisfied in the eclipse medium. Using the same parameters, the growth time of instabilities is found to be $<1$~second which will be less than the characteristic flow time for any realistic flow velocity of the medium, hence allowing instabilities in the medium to form before the medium disperses. Thus, as for PSR B1957+20 in \citet{t+94}, two of the requirements of Raman scattering are satisfied. However, although considered a possible mechanism, doubt is cast over the apparently satisfied conditions if the material is clumped, as higher densities in the medium increase the required critical flux incident on the medium. Additionally, should the medium be hot enough, the plasmon (plasma oscillations required to create turbulence) escape rate can become significant, causing a reduction in plasma turbulence present to scatter the incident radio flux. Equally, the lack of scattering observed at eclipse ingress and egress means that the transition into the eclipsing regime would have to be very abrupt. As a result, although we do not dismiss this as a possible mechanism, it appears unlikely given the observations.\\
Considered as one of the most likely eclipse mechanisms in PSRs B1957+20 \citep{t+94}, J2051$-$0827 \citep{sbl+01} and J1544+4937 \citep{brr+13} is cyclotron-synchrotron absorption. This can either occur in material entrained in the magnetised pulsar wind \citep{t+94}, or in a magnetosphere of the companion filled with relativistic particles from the pulsar wind \citep{kmg00}. Following the analysis in \citet{t+94}, to balance the pulsar wind energy density at the orbit of the companion requires a magnetic field strength, $B_{\text{E}}\sim18$~G in the eclipse region. With such a magnetic field the fundamental cyclotron frequency would be $\sim50$~MHz, thus absorption of 149~MHz and 345~MHz radiation would occur at cyclotron harmonics 3 and 7, respectively. Similar conditions calculated for PSR B1957+20 in \citet{t+94} find a cyclotron optical depth power law index at least as steep as $\tau\propto \nu^{-4}$ over the range 318--1400~MHz, for expected temperatures in the eclipse medium. Alternatively, for synchrotron absorption by relativistic electrons, the authors find a shallower frequency dependence of optical depth, $\tau\propto \nu^{-13/4}$, from their Equation (67) assuming a power law non-thermal electron distribution and magnetic field lines at an angle $45^{\circ}$ to the line-of-sight. The model in \citet{kmg00} assumes the companion to be a magnetic white dwarf, with surface magnetic field $B_{\text{S}}\gtrsim10^4$~G. The authors predict the duration of the eclipses to scale as $\propto\nu^{-0.33}$ for given temperature and electron distributions. The consistency of these models with our results suggests that cyclotron-synchrotron absorption is a comparatively likely primary eclipse mechanism for PSR J1810+1744. However a better knowledge of the eclipse medium would be required to decipher the best model.\\
\section{Conclusions}
In this paper we present a detailed account of the low frequency eclipses in the black widow pulsar system of PSR J1810+1744. We measure the dispersion, scattering and both pulsed and continuum flux densities as a function of orbital phase as a sensitive probe into the eclipsing medium. At these low frequencies we expect to be sampling the outer edges of the eclipsing medium, which we show to have a highly variable structure over timescales shorter than the 3.6~hr orbital period. The disappearance of both pulsed and continuum flux density regularly occurs between orbital phases $\sim0.18-0.35$, with little deviation in phase from one eclipse to the next. A clear frequency dependence is seen in eclipse duration, however asymmetry between ingress and egress reveal the danger of relying on duration to constrain the eclipse mechanism. Eclipse egress durations are shown to be well modelled by a single power law between 110--380~MHz, whereas the ingress shows evidence for a more complex dependence on observing frequency over this range.\\
Dispersion measure variations suggest a tail of material flows behind the companion as a result of orbital motion, similar to that seen in a number of other black widow systems. The tail material appears clumpy, and measurements show a possible hint of higher order pulse dispersion. Further high signal-to-noise observations of this bright system could attempt to track the variability of tail material and constrain a potential deviation from the typical $\propto\nu^{-2}$ dispersion relation. In addition, the inferred mass loss rate suggests that it may be possible to fully evaporate the companion in a Hubble time. Higher-frequency radio observations would probe the density of material further into the eclipse, and thus add further constraints to the inferred mass loss rate.\\
The analyses of \citet{t+94} and \citet{kmg00} are followed with the measured system parameters to add constraints on the eclipse mechanism. The eclipse duration and line-of-sight optical depth are consistent with cyclotron-synchrotron absorption as a primary eclipsing mechanism, although the apparently less likely non-linear scattering mechanisms cannot be completely ruled out. In addition, the clumpy nature of the medium could allow refraction to play a minor role. These results provide a much needed insight into the variabilities, and similarities, between eclipsing media in black widow pulsars, allowing further studies on the elusive black widow population as a whole \citep[e.g.][]{whv+17}.

\section*{Acknowledgements}
This work was based primarily on observations made with LOFAR -- the Low Frequency Array, designed and constructed by ASTRON. LOFAR has facilities in several countries, each owned and funded by various parties, that are collectively operated by the International LOFAR Telescope (ILT) foundation under a joint scientific policy. In addition, the Westerbork Synthesis Radio Telescope is operated by the ASTRON (Netherlands Institute for Radio Astronomy) with support from the Netherlands Foundation for Scientific Research (NWO). EJP would like to thank the LOFAR Pulsar Working Group, Chris Thompson, Marten van Kerkwijk and David Mulcahy for insightful comments and discussions throughout this study. This work made extensive use of \texttt{astropy} \citep{art+13} and \texttt{matplotlib} \citep{h07}. Part of this work is based on observations from the LOFAR long-term pulsar timing project (PI: J.P.W. Verbiest). EJP acknowledges support from a UK Science and Technology Facilities Council studentship. RPB acknowledges support from the ERC under the European Union's Horizon 2020 research and innovation programme (grant agreement No. 715051; Spiders). AOC gratefully acknowledges support from the European Research Council under grant ERC-2012-StG-307215 LODESTON. JWTH acknowledges funding from an NWO Vidi fellowship and from the European Research Council under the European Union's Seventh Framework Programme (FP/2007-2013) / ERC Starting Grant agreement nr. 337062 ("DRAGNET"). JvL received funding from the European Research Council under the European Union's Seventh Framework Programme (FP/2007-2013) / ERC Grant Agreement n. 617199.




\bibliographystyle{mnras}
\bibliography{../bibliography3} 

\begin{thebibliography}{}
\makeatletter
\relax
\def\mn@urlcharsother{\let\do\@makeother \do\$\do\&\do\#\do\^\do\_\do\%\do\~}
\def\mn@doi{\begingroup\mn@urlcharsother \@ifnextchar [ {\mn@doi@}
  {\mn@doi@[]}}
\def\mn@doi@[#1]#2{\def\@tempa{#1}\ifx\@tempa\@empty \href
  {http://dx.doi.org/#2} {doi:#2}\else \href {http://dx.doi.org/#2} {#1}\fi
  \endgroup}
\def\mn@eprint#1#2{\mn@eprint@#1:#2::\@nil}
\def\mn@eprint@arXiv#1{\href {http://arxiv.org/abs/#1} {{\tt arXiv:#1}}}
\def\mn@eprint@dblp#1{\href {http://dblp.uni-trier.de/rec/bibtex/#1.xml}
  {dblp:#1}}
\def\mn@eprint@#1:#2:#3:#4\@nil{\def\@tempa {#1}\def\@tempb {#2}\def\@tempc
  {#3}\ifx \@tempc \@empty \let \@tempc \@tempb \let \@tempb \@tempa \fi \ifx
  \@tempb \@empty \def\@tempb {arXiv}\fi \@ifundefined
  {mn@eprint@\@tempb}{\@tempb:\@tempc}{\expandafter \expandafter \csname
  mn@eprint@\@tempb\endcsname \expandafter{\@tempc}}}

\bibitem[\protect\citeauthoryear{{Akaike}}{{Akaike}}{1974}]{a74}
{Akaike} H.,  1974, IEEE Transactions on Automatic Control, \href
  {http://adsabs.harvard.edu/abs/1974ITAC...19..716A} {19, 716}

\bibitem[\protect\citeauthoryear{{Alexov}, {Hessels}, {Mol}, {Stappers}  \&
  {van Leeuwen}}{{Alexov} et~al.}{2010}]{ahm+10}
{Alexov} A.,  {Hessels} J.,  {Mol} J.~D.,  {Stappers} B.,   {van Leeuwen} J.,
  2010, in {Mizumoto} Y.,  {Morita} K.-I.,   {Ohishi} M.,  eds,  Astronomical
  Society of the Pacific Conference Series Vol. 434, Astronomical Data Analysis
  Software and Systems XIX. p.~193 (\mn@eprint {arXiv} {1012.1583})

\bibitem[\protect\citeauthoryear{{Archibald} et~al.,}{{Archibald}
  et~al.}{2009}]{asr+09}
{Archibald} A.~M.,  et~al., 2009, Science, 324, 1411

\bibitem[\protect\citeauthoryear{{Archibald}, {Kondratiev}, {Hessels}  \&
  {Stinebring}}{{Archibald} et~al.}{2014}]{akh+14}
{Archibald} A.~M.,  {Kondratiev} V.~I.,  {Hessels} J.~W.~T.,   {Stinebring}
  D.~R.,  2014, \mn@doi [\apjl] {10.1088/2041-8205/790/2/L22}, \href
  {http://adsabs.harvard.edu/abs/2014ApJ...790L..22A} {790, L22}

\bibitem[\protect\citeauthoryear{{Arzoumanian}, {Fruchter}  \&
  {Taylor}}{{Arzoumanian} et~al.}{1994}]{aft94}
{Arzoumanian} Z.,  {Fruchter} A.~S.,   {Taylor} J.~H.,  1994, The Astrophysical
  Journal, Letters, 426, L85

\bibitem[\protect\citeauthoryear{{Astropy Collaboration} et~al.,}{{Astropy
  Collaboration} et~al.}{2013}]{art+13}
{Astropy Collaboration} et~al., 2013, Astronomy and Astrophysics, 558

\bibitem[\protect\citeauthoryear{{Baars} \& {Hooghoudt}}{{Baars} \&
  {Hooghoudt}}{1974}]{bh74}
{Baars} J.~W.~M.,  {Hooghoudt} B.~G.,  1974, \aap, \href
  {http://adsabs.harvard.edu/abs/1974A%26A....31..323B} {31, 323}

\bibitem[\protect\citeauthoryear{{Bhattacharyya} et~al.,}{{Bhattacharyya}
  et~al.}{2013}]{brr+13}
{Bhattacharyya} B.,  et~al., 2013, The Astrophysical Journal, Letters, 773, L12

\bibitem[\protect\citeauthoryear{{Bilous} et~al.,}{{Bilous}
  et~al.}{2016}]{bkk+16}
{Bilous} A.~V.,  et~al., 2016, \mn@doi [\aap] {10.1051/0004-6361/201527702},
  \href {http://adsabs.harvard.edu/abs/2016A%26A...591A.134B} {591, A134}

\bibitem[\protect\citeauthoryear{{Breton} et~al.,}{{Breton}
  et~al.}{2013}]{bvr+13}
{Breton} R.~P.,  et~al., 2013, The Astrophysical Journal, 769, 108

\bibitem[\protect\citeauthoryear{{Broderick} et~al.,}{{Broderick}
  et~al.}{2016}]{bfb+16}
{Broderick} J.~W.,  et~al., 2016, \mn@doi [\mnras] {10.1093/mnras/stw794},
  \href {http://adsabs.harvard.edu/abs/2016MNRAS.459.2681B} {459, 2681}

\bibitem[\protect\citeauthoryear{{Camilo} et~al.,}{{Camilo}
  et~al.}{2015}]{ckr+15}
{Camilo} F.,  et~al., 2015, \mn@doi [\apj] {10.1088/0004-637X/810/2/85}, \href
  {http://adsabs.harvard.edu/abs/2015ApJ...810...85C} {810, 85}

\bibitem[\protect\citeauthoryear{{Cromartie} et~al.,}{{Cromartie}
  et~al.}{2016}]{cck+16}
{Cromartie} H.~T.,  et~al., 2016, \mn@doi [\apj] {10.3847/0004-637X/819/1/34},
  \href {http://adsabs.harvard.edu/abs/2016ApJ...819...34C} {819, 34}

\bibitem[\protect\citeauthoryear{{Eichler} \& {Levinson}}{{Eichler} \&
  {Levinson}}{1988}]{el88}
{Eichler} D.,  {Levinson} A.,  1988, \mn@doi [\apjl] {10.1086/185341}, \href
  {http://adsabs.harvard.edu/abs/1988ApJ...335L..67E} {335, L67}

\bibitem[\protect\citeauthoryear{{Emmering} \& {London}}{{Emmering} \&
  {London}}{1990}]{el90}
{Emmering} R.~T.,  {London} R.~A.,  1990, The Astrophysical Journal, 363, 589

\bibitem[\protect\citeauthoryear{{Fruchter}, {Stinebring}  \&
  {Taylor}}{{Fruchter} et~al.}{1988a}]{fst88}
{Fruchter} A.~S.,  {Stinebring} D.~R.,   {Taylor} J.~H.,  1988a, Nature, 333,
  237

\bibitem[\protect\citeauthoryear{{Fruchter}, {Gunn}, {Lauer}  \&
  {Dressler}}{{Fruchter} et~al.}{1988b}]{f+88}
{Fruchter} A.~S.,  {Gunn} J.~E.,  {Lauer} T.~R.,   {Dressler} A.,  1988b,
  Nature, 334, 686

\bibitem[\protect\citeauthoryear{{Fruchter} et~al.,}{{Fruchter}
  et~al.}{1990}]{fbb+90}
{Fruchter} A.~S.,  et~al., 1990, The Astrophysical Journal, 351, 642

\bibitem[\protect\citeauthoryear{{Gentile} et~al.,}{{Gentile}
  et~al.}{2013}]{gmr+13}
{Gentile} P.,  et~al., 2013, in {van Leeuwen} J.,  ed.,  IAU Symposium Vol.
  291, IAU Symposium. pp 389--391 (\mn@eprint {arXiv} {1210.7342})

\bibitem[\protect\citeauthoryear{{Geyer} et~al.,}{{Geyer}
  et~al.}{2017}]{gkk+17}
{Geyer} M.,  et~al., 2017, preprint, \href
  {http://adsabs.harvard.edu/abs/2017arXiv170604205G} {} (\mn@eprint {arXiv}
  {1706.04205})

\bibitem[\protect\citeauthoryear{{Hamaker}}{{Hamaker}}{2006}]{h06}
{Hamaker} J.~P.,  2006, Astronomy and Astrophysics, 456, 395

\bibitem[\protect\citeauthoryear{{Haslam}, {Salter}, {Stoffel}  \&
  {Wilson}}{{Haslam} et~al.}{1982}]{hss+82}
{Haslam} C.~G.~T.,  {Salter} C.~J.,  {Stoffel} H.,   {Wilson} W.~E.,  1982,
  Astronomy and Astrophysics, Supplement, 47, 1

\bibitem[\protect\citeauthoryear{{Hassall} et~al.,}{{Hassall}
  et~al.}{2012}]{hsh+12}
{Hassall} T.~E.,  et~al., 2012, \mn@doi [\aap] {10.1051/0004-6361/201218970},
  \href {http://adsabs.harvard.edu/abs/2012A%26A...543A..66H} {543, A66}

\bibitem[\protect\citeauthoryear{{Heald} et~al.,}{{Heald}
  et~al.}{2010}]{hmp+10}
{Heald} G.,  et~al., 2010, preprint, \href
  {http://adsabs.harvard.edu/abs/2010arXiv1008.4693H} {} (\mn@eprint {arXiv}
  {1008.4693})

\bibitem[\protect\citeauthoryear{{Hessels} et~al.,}{{Hessels}
  et~al.}{2011}]{hrm+11}
{Hessels} J.~W.~T.,  et~al., 2011, in {Burgay} M.,  {D'Amico} N.,  {Esposito}
  P.,  {Pellizzoni} A.,   {Possenti} A.,  eds,  American Institute of Physics
  Conference Series Vol. 1357, American Institute of Physics Conference Series.
  pp 40--43 (\mn@eprint {arXiv} {1101.1742})

\bibitem[\protect\citeauthoryear{{Hobbs}, {Edwards}  \& {Manchester}}{{Hobbs}
  et~al.}{2006}]{hem06}
{Hobbs} G.,  {Edwards} R.,   {Manchester} R.,  2006, Chinese Journal of
  Astronomy and Astrophysics Supplement, \href
  {http://adsabs.harvard.edu/abs/2006ChJAS...6b.189H} {6, 189}

\bibitem[\protect\citeauthoryear{{Hotan}, {van Straten}  \&
  {Manchester}}{{Hotan} et~al.}{2004}]{hvm04}
{Hotan} A.~W.,  {van Straten} W.,   {Manchester} R.~N.,  2004, Publications of
  the Astron. Soc. of Australia, 21, 302

\bibitem[\protect\citeauthoryear{{Hunter}}{{Hunter}}{2007}]{h07}
{Hunter} J.~D.,  2007, \mn@doi [Computing in Science and Engineering]
  {10.1109/MCSE.2007.55}, \href
  {http://adsabs.harvard.edu/abs/2007CSE.....9...90H} {9, 90}

\bibitem[\protect\citeauthoryear{{Hurley-Walker} et~al.,}{{Hurley-Walker}
  et~al.}{2017}]{hch+17}
{Hurley-Walker} N.,  et~al., 2017, \mn@doi [\mnras] {10.1093/mnras/stw2337},
  \href {http://adsabs.harvard.edu/abs/2017MNRAS.464.1146H} {464, 1146}

\bibitem[\protect\citeauthoryear{{Intema}, {Jagannathan}, {Mooley}  \&
  {Frail}}{{Intema} et~al.}{2017}]{ijm+17}
{Intema} H.~T.,  {Jagannathan} P.,  {Mooley} K.~P.,   {Frail} D.~A.,  2017,
  \mn@doi [\aap] {10.1051/0004-6361/201628536}, \href
  {http://adsabs.harvard.edu/abs/2017A%26A...598A..78I} {598, A78}

\bibitem[\protect\citeauthoryear{{Karuppusamy}, {Stappers}  \& {van
  Straten}}{{Karuppusamy} et~al.}{2008}]{ksv08}
{Karuppusamy} R.,  {Stappers} B.,   {van Straten} W.,  2008, \mn@doi [\pasp]
  {10.1086/528699}, \href {http://adsabs.harvard.edu/abs/2008PASP..120..191K}
  {120, 191}

\bibitem[\protect\citeauthoryear{{Khechinashvili}, {Melikidze}  \&
  {Gil}}{{Khechinashvili} et~al.}{2000}]{kmg00}
{Khechinashvili} D.~G.,  {Melikidze} G.~I.,   {Gil} J.~A.,  2000, The
  Astrophysical Journal, 541, 335

\bibitem[\protect\citeauthoryear{{Kondratiev} et~al.,}{{Kondratiev}
  et~al.}{2016}]{kvh+16}
{Kondratiev} V.~I.,  et~al., 2016, \aap, 585, A128

\bibitem[\protect\citeauthoryear{{Kramer}, {Lange}, {Lorimer}, {Backer},
  {Xilouris}, {Jessner}  \& {Wielebinski}}{{Kramer} et~al.}{1999}]{k+99}
{Kramer} M.,  {Lange} C.,  {Lorimer} D.~R.,  {Backer} D.~C.,  {Xilouris} K.~M.,
   {Jessner} A.,   {Wielebinski} R.,  1999, The Astrophysical Journal, 526, 957

\bibitem[\protect\citeauthoryear{{Kuniyoshi}, {Verbiest}, {Lee}, {Adebahr},
  {Kramer}  \& {Noutsos}}{{Kuniyoshi} et~al.}{2015}]{kvl+15}
{Kuniyoshi} M.,  {Verbiest} J.~P.~W.,  {Lee} K.~J.,  {Adebahr} B.,  {Kramer}
  M.,   {Noutsos} A.,  2015, Monthly Notices of the Royal Astronomical Society,
  453, 828

\bibitem[\protect\citeauthoryear{{Kurtz}}{{Kurtz}}{2002}]{k02}
{Kurtz} S.,  2002, in {Crowther} P.,  ed.,  Astronomical Society of the Pacific
  Conference Series Vol. 267, Hot Star Workshop III: The Earliest Phases of
  Massive Star Birth. p.~81 (\mn@eprint {} {astro-ph/0111351})

\bibitem[\protect\citeauthoryear{{Kuzmin} \& {Losovsky}}{{Kuzmin} \&
  {Losovsky}}{2001}]{kl01}
{Kuzmin} A.~D.,  {Losovsky} B.~Y.,  2001, \mn@doi [\aap]
  {10.1051/0004-6361:20000507}, \href
  {http://adsabs.harvard.edu/abs/2001A%26A...368..230K} {368, 230}

\bibitem[\protect\citeauthoryear{{Lang}}{{Lang}}{1971}]{l71}
{Lang} K.~R.,  1971, \apj, 164, 249

\bibitem[\protect\citeauthoryear{{Lawson}, {Mayer}, {Osborne}  \&
  {Parkinson}}{{Lawson} et~al.}{1987}]{lmo+87}
{Lawson} K.~D.,  {Mayer} C.~J.,  {Osborne} J.~L.,   {Parkinson} M.~L.,  1987,
  Monthly Notices of the Royal Astronomical Society, 225, 307

\bibitem[\protect\citeauthoryear{{Lee} \& {Jokipii}}{{Lee} \&
  {Jokipii}}{1976}]{lj76}
{Lee} L.~C.,  {Jokipii} J.~R.,  1976, \apj, 206, 735

\bibitem[\protect\citeauthoryear{{Lyne}, {Johnston}, {Manchester},
  {Staveley-Smith}  \& {D'Amico}}{{Lyne} et~al.}{1990}]{ljm+90}
{Lyne} A.~G.,  {Johnston} S.,  {Manchester} R.~N.,  {Staveley-Smith} L.,
  {D'Amico} N.,  1990, Nature, 347, 650

\bibitem[\protect\citeauthoryear{{Murphy} et~al.,}{{Murphy}
  et~al.}{2017}]{mkb+17}
{Murphy} T.,  et~al., 2017, \mn@doi [\pasa] {10.1017/pasa.2017.13}, \href
  {http://adsabs.harvard.edu/abs/2017PASA...34...20M} {34, e020}

\bibitem[\protect\citeauthoryear{{Noutsos} et~al.,}{{Noutsos}
  et~al.}{2015}]{nsk+15}
{Noutsos} A.,  et~al., 2015, \aap, 576, A62

\bibitem[\protect\citeauthoryear{{Offringa}, {de Bruyn}, {Biehl}, {Zaroubi},
  {Bernardi}  \& {Pandey}}{{Offringa} et~al.}{2010}]{odb+10}
{Offringa} A.~R.,  {de Bruyn} A.~G.,  {Biehl} M.,  {Zaroubi} S.,  {Bernardi}
  G.,   {Pandey} V.~N.,  2010, \mnras, 405, 155

\bibitem[\protect\citeauthoryear{{Offringa}, {de Bruyn}  \&
  {Zaroubi}}{{Offringa} et~al.}{2012a}]{odz12}
{Offringa} A.~R.,  {de Bruyn} A.~G.,   {Zaroubi} S.,  2012a, \mn@doi [\mnras]
  {10.1111/j.1365-2966.2012.20633.x}, \href
  {http://adsabs.harvard.edu/abs/2012MNRAS.422..563O} {422, 563}

\bibitem[\protect\citeauthoryear{{Offringa}, {van de Gronde}  \&
  {Roerdink}}{{Offringa} et~al.}{2012b}]{ovr12}
{Offringa} A.~R.,  {van de Gronde} J.~J.,   {Roerdink} J.~B.~T.~M.,  2012b,
  \mn@doi [\aap] {10.1051/0004-6361/201118497}, \href
  {http://adsabs.harvard.edu/abs/2012A%26A...539A..95O} {539, A95}

\bibitem[\protect\citeauthoryear{{Pandey}, {van Zwieten}, {de Bruyn}  \&
  {Nijboer}}{{Pandey} et~al.}{2009}]{pvd+09}
{Pandey} V.~N.,  {van Zwieten} J.~E.,  {de Bruyn} A.~G.,   {Nijboer} R.,  2009,
  in {Saikia} D.~J.,  {Green} D.~A.,  {Gupta} Y.,   {Venturi} T.,  eds,
  Astronomical Society of the Pacific Conference Series Vol. 407, The
  Low-Frequency Radio Universe. p.~384

\bibitem[\protect\citeauthoryear{{Phillips} \& {Wolszczan}}{{Phillips} \&
  {Wolszczan}}{1992}]{pw92}
{Phillips} J.~A.,  {Wolszczan} A.,  1992, \mn@doi [\apj] {10.1086/170935},
  \href {http://adsabs.harvard.edu/abs/1992ApJ...385..273P} {385, 273}

\bibitem[\protect\citeauthoryear{{Ransom}}{{Ransom}}{2001}]{r01}
{Ransom} S.~M.,  2001, in American Astronomical Society Meeting Abstracts.
  p.~119

\bibitem[\protect\citeauthoryear{{Ray} et~al.,}{{Ray} et~al.}{2012}]{rap+12}
{Ray} P.~S.,  et~al., 2012, preprint (\mn@eprint {arXiv} {1205.3089})

\bibitem[\protect\citeauthoryear{{Roberts}}{{Roberts}}{2013}]{r13}
{Roberts} M.~S.~E.,  2013, in {van Leeuwen} J.,  ed.,  IAU Symposium Vol. 291,
  IAU Symposium. pp 127--132 (\mn@eprint {arXiv} {1210.6903})

\bibitem[\protect\citeauthoryear{{Ruderman}, {Shaham}  \& {Tavani}}{{Ruderman}
  et~al.}{1989}]{rst89rud}
{Ruderman} M.,  {Shaham} J.,   {Tavani} M.,  1989, \mn@doi [\apj]
  {10.1086/167029}, \href {http://adsabs.harvard.edu/abs/1989ApJ...336..507R}
  {336, 507}

\bibitem[\protect\citeauthoryear{{Ryba} \& {Taylor}}{{Ryba} \&
  {Taylor}}{1991}]{rt91}
{Ryba} M.~F.,  {Taylor} J.~H.,  1991, \mn@doi [\apj] {10.1086/170613}, \href
  {http://adsabs.harvard.edu/abs/1991ApJ...380..557R} {380, 557}

\bibitem[\protect\citeauthoryear{{Savitzky} \& {Golay}}{{Savitzky} \&
  {Golay}}{1964}]{sg64}
{Savitzky} A.,  {Golay} M.~J.~E.,  1964, Analytical Chemistry, \href
  {http://adsabs.harvard.edu/abs/1964AnaCh..36.1627S} {36, 1627}

\bibitem[\protect\citeauthoryear{{Scaife} \& {Heald}}{{Scaife} \&
  {Heald}}{2012}]{sh12}
{Scaife} A.~M.~M.,  {Heald} G.~H.,  2012, \mn@doi [\mnras]
  {10.1111/j.1745-3933.2012.01251.x}, \href
  {http://adsabs.harvard.edu/abs/2012MNRAS.423L..30S} {423, L30}

\bibitem[\protect\citeauthoryear{{Schroeder} \& {Halpern}}{{Schroeder} \&
  {Halpern}}{2014}]{sh14}
{Schroeder} J.,  {Halpern} J.,  2014, The Astrophysical Journal, 793, 78

\bibitem[\protect\citeauthoryear{{Schwarz}}{{Schwarz}}{1978}]{s78}
{Schwarz} G.,  1978, Annals of Statistics, \href
  {http://adsabs.harvard.edu/abs/1978AnSta...6..461S} {6, 461}

\bibitem[\protect\citeauthoryear{{Shaifullah} et~al.,}{{Shaifullah}
  et~al.}{2016}]{svf+16}
{Shaifullah} G.,  et~al., 2016, \mn@doi [\mnras] {10.1093/mnras/stw1737}, \href
  {http://adsabs.harvard.edu/abs/2016MNRAS.462.1029S} {462, 1029}

\bibitem[\protect\citeauthoryear{{Stappers} et~al.,}{{Stappers}
  et~al.}{1996}]{s+96}
{Stappers} B.~W.,  et~al., 1996, The Astrophysical Journal, Letters, 465, L119

\bibitem[\protect\citeauthoryear{{Stappers}, {Bailes}, {Lyne}, {Camilo},
  {Manchester}, {Sandhu}, {Toscano}  \& {Bell}}{{Stappers}
  et~al.}{2001}]{sbl+01}
{Stappers} B.~W.,  {Bailes} M.,  {Lyne} A.~G.,  {Camilo} F.,  {Manchester}
  R.~N.,  {Sandhu} J.~S.,  {Toscano} M.,   {Bell} J.~F.,  2001, Monthly Notices
  of the Royal Astronomical Society, 321, 576

\bibitem[\protect\citeauthoryear{{Stappers} et~al.,}{{Stappers}
  et~al.}{2011}]{s+11}
{Stappers} B.~W.,  et~al., 2011, Astronomy and Astrophysics, 530, A80

\bibitem[\protect\citeauthoryear{{Thompson}, {Blandford}, {Evans}  \&
  {Phinney}}{{Thompson} et~al.}{1994}]{t+94}
{Thompson} C.,  {Blandford} R.~D.,  {Evans} C.~R.,   {Phinney} E.~S.,  1994,
  The Astrophysical Journal, 422, 304

\bibitem[\protect\citeauthoryear{{Wadiasingh}, {Harding}, {Venter},
  {B{\"o}ttcher}  \& {Baring}}{{Wadiasingh} et~al.}{2017}]{whv+17}
{Wadiasingh} Z.,  {Harding} A.~K.,  {Venter} C.,  {B{\"o}ttcher} M.,   {Baring}
  M.~G.,  2017, \mn@doi [\apj] {10.3847/1538-4357/aa69bf}, \href
  {http://adsabs.harvard.edu/abs/2017ApJ...839...80W} {839, 80}

\bibitem[\protect\citeauthoryear{{Wijnholds} \& {van Cappellen}}{{Wijnholds} \&
  {van Cappellen}}{2011}]{wv11}
{Wijnholds} S.~J.,  {van Cappellen} W.~A.,  2011, IEEE Transactions on Antennas
  and Propagation, 59, 1981

\bibitem[\protect\citeauthoryear{{Williamson}}{{Williamson}}{1972}]{w72}
{Williamson} I.~P.,  1972, \mn@doi [\mnras] {10.1093/mnras/157.1.55}, \href
  {http://adsabs.harvard.edu/abs/1972MNRAS.157...55W} {157, 55}

\bibitem[\protect\citeauthoryear{{van Haarlem} et~al.,}{{van Haarlem}
  et~al.}{2013}]{v+13}
{van Haarlem} M.~P.,  et~al., 2013, Astronomy and Astrophysics, 556, A2

\bibitem[\protect\citeauthoryear{{van Leeuwen} \& {Stappers}}{{van Leeuwen} \&
  {Stappers}}{2010}]{vs10}
{van Leeuwen} J.,  {Stappers} B.~W.,  2010, \mn@doi [\aap]
  {10.1051/0004-6361/200913121}, \href
  {http://adsabs.harvard.edu/abs/2010A%26A...509A...7V} {509, A7}

\bibitem[\protect\citeauthoryear{{van Straten} \& {Bailes}}{{van Straten} \&
  {Bailes}}{2011}]{vb11}
{van Straten} W.,  {Bailes} M.,  2011, Publications of the Astron. Soc. of
  Australia, 28, 1

\makeatother
\end{thebibliography}




\appendix

\section{Ephemerides}\label{Sec: orb_ephem}
Table~\ref{Table: ephem1} shows the parameters used to fold the beamformed observations of PSR J1810+1744 in this work. Due to unsatisfactory long-term solutions, adjusted ephemerides were used for most observations. The adjusted parameters, and their modified values are shown in Table~\ref{Table: ephem2}.\\

\begin{table}
	\centering
	\caption{Parameters for PSR J1810+1744.}
	\label{Table: ephem1}
	\begin{tabular}{ll}
		\hline
		Parameter & Value \\
		\hline
		RAJ & 18:10:37.2817102 \\
		DECJ & +17:44:37.36758 \\  
		F0 & 601.41150960535386139 \\
		F1 & -1.6320154486270032265e-15 \\
		PEPOCH & 56043.457048008625684       \\
		POSEPOCH & 55530.000192047375631       \\
		DMEPOCH & 56043.456848000000001       \\
		DM & 39.657544844694188023       \\
		PMRA & 18.478282659525929557  \\
		PMDEC & 2.5348579936231234319 \\
		BINARY & BT \\
		PB & 0.14817027505277922673 \\   
		T0 & 55130.048390573342001 \\   
		A1 & 0.095385516379427031886 \\   
		OM & 0 \\
		ECC & 0 \\
		START & 57071.294732713806226 \\
		FINISH & 57073.494745636093214 \\
		\hline
	\end{tabular}
\end{table}

\begin{table*}
	\centering
	\caption{List of observations and the corresponding adjusted ephemeris parameters. $^a$L260707, L260713, L260719, L260725, L260731, L260737, L260743 and L260749.}
	\label{Table: ephem2}
	\begin{tabular}{lccccl}
		\hline	
		Date & Telescope & Project ID & ObsID & Duration & Adjusted parameters \\
		\hline
		$2015 / 02 / 19$ & LOFAR & LC2\_039 & $^a$ & $8\times30$m & As above \\
		\hline
		$2014 / 09 / 19$ & LOFAR & LC2\_026 & L243355 & 1h & F0 = 601.4114975594 \\
		& & & & & START = 56824.005317681616837 \\
		& & & & & FINISH = 56826.046924821248943 \\
		\hline
		$2014 / 09 / 18$ & LOFAR & LC2\_026 & L243365 & 1h & F0 = 601.41149755942529537 \\
		& & & & & START = 56824.005317681616837 \\
		& & & & & FINISH = 56826.046924821248943 \\
		\hline
		$2014 / 06 / 17$ & LOFAR & LC2\_026 & L231759 & 1h & F0 = 601.4114975594 \\
		& & & & & START = 56824.005317681616837 \\
		& & & & & FINISH = 56826.046924821248943 \\
		\hline
		$2013 / 10 / 02$ & LOFAR & LC0\_011 & L181912 & 5m & F0 = 601.4114975594 \\
		& & & & & START = 56824.005317681616837 \\
		& & & & & FINISH = 56826.046924821248943 \\
		\hline
		$2013 / 07 / 26$ & LOFAR & DDT\_005 & L166106 & 2h 6m & F0 = 601.4115075853279595 \\
		& & & & & F1 = $-7.0431312170082538772 \times 10^{-14}$ \\
		& & & & & T0 = 55130.048424058501556 \\
		& & & & & START = 56496.838153045821123 \\
		& & & & & FINISH = 56498.931146729934149 \\
		\hline
		$2013 / 07 / 25$ & LOFAR & DDT\_005 & L166110 & 2h 42m & F0 = 601.4115075853279595 \\
		& & & & & F1 = $1.422881594092733325 \times 10^{-14}$ \\
		& & & & & T0 = 55130.048422567561047 \\
		& & & & & START = 56497.817284424338311 \\
		& & & & & FINISH = 56499.92833244848606 \\
		\hline
		$2013 / 07 / 24$ & LOFAR & DDT\_005 & L165450 & 2h 12m & F0 = 601.4115075853279595 \\
		& & & & & F1 = $1.422881594092733325 \times 10^{-14}$ \\
		& & & & & T0 = 55130.048422567561047 \\
		& & & & & START = 56497.817284424338311 \\
		& & & & & FINISH = 56499.92833244848606 \\
		\hline
		$2012 / 12 / 20$ & LOFAR & LC0\_011 & L81280 & 20m & F0 = 601.41147769170566595 \\
		& & & & & START = 56281.496642146127428 \\
		& & & & & FINISH = 56281.51149934484264 \\
		\hline
		$2011 / 06 / 06$ & WSRT & S11A008 & 11102762 & 5h & F0 = 601.4115075853279595 \\
		& & & & & F1 = $3.043269057704096029 \times 10^{-14}$ \\
		& & & & & T0 = 55130.04835922778463 \\
		& & & & & A1 = 0.095377871058077957022 \\
		& & & & & START = 55717.076542729752688 \\
		& & & & & FINISH = 55719.284128548818259 \\
		\hline
	\end{tabular}
\end{table*}


\bsp	
\label{lastpage}
\end{document}